\documentclass[prb,twocolumn,amsmath,amssymb,showpacs,superscriptaddress,floatfix]{revtex4}

\usepackage{epsfig}
\usepackage{fancyhdr}
\usepackage{amssymb} 
\usepackage{lscape}
\usepackage{dcolumn}

\newcommand{\csixty}{C$_{60}$}
\newcommand{\csixtyminus}{C$_{60}^-$}
\newcommand{\csixtytwominus}{C$_{60}^{2-}$}
\newcommand{\csixtythreeminus}{C$_{60}^{3-}$}
\newcommand{\csixtyfourminus}{C$_{60}^{4-}$}

\newcommand{\csixtynminus}{C$_{60}^{N-}$}

\newcommand{\kfourcsixty}{K$_4$C$_{60}$}

\newcommand{\athreecsixty}{A$_3$C$_{60}$}

\newcommand{\tl}[1]{_{\textrm{#1}}}
\newcommand{\tu}[1]{^{\textrm{#1}}}
\newcommand{\V}[1]{{\bf #1}}
\newcommand{\ket}[1]{|#1\rangle}
\newcommand{\bra}[1]{\langle#1|}

\begin{document}

\title{Jahn-Teller effect versus Hund's rule coupling in C$_{60}^{N-}$}

\author{S. Wehrli}
\email{samuel.wehrli@alumni.ethz}
\author{M. Sigrist}
\affiliation{ Theoretische Physik, ETH-H\"onggerberg, CH-8093 Z\"urich, Switzerland}

\date{\today}

\begin{abstract}
We propose variational states for the ground state and the low-energy collective rotator excitations in negatively charged C$_{60}^{N-}$ ions ($N=1\ldots 5$). The approach includes the linear electron-phonon coupling and the Coulomb interaction on the same level. The electron-phonon coupling is treated within the effective mode approximation (EMA) which yields the linear $t_{1u}\otimes H_g$ Jahn-Teller problem whereas the Coulomb interaction gives rise to Hund's rule coupling for $N=2,3,4$. The Hamiltonian has accidental SO(3) symmetry which allows an elegant formulation in terms of angular momenta. Trial states are constructed from coherent states and using projection operators onto angular momentum subspaces which results in good variational states for the complete parameter range. The evaluation of the corresponding energies is to a large extent analytical. We use the approach for a detailed analysis of the competition between Jahn-Teller effect and Hund's rule coupling, which determines the spin state for $N=2,3,4$. We calculate the low-spin/high-spin gap for $N=2,3,4$ as a function of the Hund's rule coupling constant $J$. We find that the experimentally measured gaps suggest a coupling constant in the range $J=60-80$~meV. Using a finite value for $J$, we recalculate the ground state energies of the C$_{60}^{N-}$ ions and find that the Jahn-Teller energy gain is partly counterbalanced by the Hund's rule coupling. In particular, the ground state energies for $N=2,3,4$ are almost equal.
\end{abstract}


\maketitle

\section{Introduction}\label{sec:intro}

The $t_{1u}\otimes H_g$ Jahn-Teller problem, where electrons in a threefold degenerate orbital interact with a fivefold degenerate phonon multiplet, is known since more than 30~years. It first arised for the particular case of $p$-electrons in a cubic systems which are equally coupled to $E_g$ and $T_{2g}$ vibrational modes~\cite{art:obrien69}. On the level of linear coupling, an equivalent problem arises in negatively charged \csixtynminus{} ions ($N=1\ldots5$). These materials experienced particular interest when superconductivity was observed in alkali-doped \athreecsixty{} (A=K,Cs,Rb, for a review see Ref.~\onlinecite{art:gunnarsson97}). The neutral \csixty{} molecule is a closed shell system and highly symmetric. It has icosahedral symmetry which is the largest threedimensional point-group with 1-, 3-, 4- and 5-dimensional irreducible representations (IR)~\cite{book:Butler}. The lowest unoccupied molecular orbital (LUMO) of \csixty{} is threefold degenerate and has $t_{1u}$ symmetry. It couples to two non-degenerate $A_g$ phonon modes and eight 5-fold degenerate $H_g$ phonon multiplets~\cite{art:lannoo91,art:varma91}. In the present work we focus on the non-trivial coupling to the $H_g$ multiplets. We restrict our attention to linear coupling and approximate the eight $H_g$ multiplets by one effective multiplet which gives rise to the linear, single-mode $t_{1u}\otimes H_g$ Jahn-Teller problem. Furthermore, we will use the fact that the icosahedral IR's $t_{1u}$ and $H_g$ correspond to the $L=1,2$ IR of SO(3) which don't split under the icosahedral symmetry~\cite{book:Butler}. As a consequence, the linear $t_{1u}\otimes H_g$ Jahn-Teller problem is equivalent to the problem of $p$-electrons interacting with $d$-phonons and recovers accidental SO(3) symmetry. Therefore, the present work really treats the linear $p\otimes d$ Jahn-Teller problem and we will mostly speak of $p$-electrons and $d$-phonons throughout the following. Note that the accidental SO(3) symmetry would be lifted by higher order coupling terms~\cite{art:dunn95,book:Chancey}. In addition to electron-phonon coupling, electrons in \csixty{} also interact via the Coulomb interaction. This leads to the so-called Hund's rule coupling (see Ref.~\onlinecite{art:wierzbowska04} and references therein). Below, we will consider both, the Jahn-Teller and Hund's rule interaction, and discuss the competition between them.

The $p\otimes d$ Jahn-Teller cannot be trivially solved as, for example, a single displaced harmonic oscillator. This led to various different approaches. Early work on the case $N=1$ was done by O'Brien~\cite{art:obrien69,art:obrien71,art:obrien72}. The first treatment for all fillings was carried out in two subsequent papers by Auerbach, Manini and Tosatti~\cite{art:auerbach94,art:manini94}. The first paper is based on a semiclassical (also called adiabatic) approximation which yields the effective Hamiltonian in the strong coupling limit~\cite{art:auerbach94}. The intermediate regime is explored using exact diagonalization. The weak coupling limit is treated in the second paper using perturbation theory~\cite{art:manini94}. The effect of Hund's-rule coupling was studied subsequently using the same approach~\cite{art:obrien96}. These works led to a good understanding of the $t_{1u}\otimes H_g$ Jahn-Teller problem. A complete discussion is given in Ref.~\onlinecite{book:Chancey}.

However, none of the schemes just discussed applies to the whole coupling range. Even ``exact'' diagonalization is only valid for the small and intermediate regime because it suffers from truncation of the phonon Hilbert space at high coupling. Moreover, the approaches don't provide wavefunctions in the intermediate regime, which is precisely the regime of \csixty{}. There have been various attempts to construct wavefunctions for all coupling regimes on the basis of coherent states and projection techniques~\cite{art:judd75,art:chancey87,art:dunn02,art:sookhun03,art:dunn05}. The formalisms used were rather involved and led to complicated analytical expressions and multidimensional integrals. Here, we propose a mathematically equivalent but much more convenient formalism which is based on the use of projection operators. 

We construct variational states in two steps: First, we start with a product state $\ket{\Psi(q)}=\ket \psi_e\otimes \ket{q}_p$ which minimizes the electron-phonon coupling. The phonon part $\ket{q}_p$ is a coherent state which corresponds to the displacement $q$ of one phonon coordinate. This displacement leads to a splitting of the degenerate electronic levels. The electron part $\ket \psi_e$ is chosen such as to minimize the energy of the electrons for a given splitting. The state $\ket\Psi$ is not SO(3) symmetric and also not a good angular momentum state. However, the SO(3) symmetry of the Hamiltonian requires eigenstates to be angular momentum states. Therefore, we use projection operators $Q^L_{MK}$, as defined in~(\ref{eq:QLMK}), to construct angular momentum states $\ket{LMK,q}=Q^L_{MK}\ket{\Psi(q)}$. State $\ket{LMK,q}$ is the final variational wavefunction with good quantum numbers $LM$ and one variational parameter $q$. The expectation value of $H$ with respect to this state is
$\bra{\Psi}Q^L_{KM}HQ^L_{MK}\ket{\psi}=\bra{\Psi}HQ^L_{KK}\ket{\psi}$. The equality arises because $H$ is a scalar and commutes with the projection operator. This is an essential simplification because each projection operator carries an integration over Euler angles, as can be seen from definition~(\ref{eq:QLMK}). Previous works haven't made use of this property which, as we show below, allows an almost analytical treatment of the problem.

The present approach is interesting because it is to a large extent analytical and applies to the whole coupling range. Energies can be calculated with moderate effort which allows a detailed analysis of the competition between Jahn-Teller effect and Hund's rule coupling. This competition determines the spin configuration of the ground state for the cases $N=2,3,4$ which we calculate for the complete parameter range. 

The paper is organized as follows: In section~\ref{sec:ham} we introduce the Hamiltonian as well as the effective mode approximation which reduces the multi-mode problem to a single-mode problem. The properties of the phonon coherent states are discussed in section~\ref{sec:coh}. Section~\ref{sec:c60m1} to~\ref{sec:c60m3} treat the cases $N=1,2,3$ respectively. As discussed below, the cases $N=4,5$ are equivalent to $N=2,1$ due to particle-hole symmetry. Results specific to parameters of \csixty{} are discussed in section~\ref{sec:c60spec}. Due to the SO(3) symmetry, angular momenta and its eigenstates, the spherical harmonics, play an important role in the present work. We always use real spherical harmonics $Y_{LM}(\theta,\phi)$ which have $\cos(M\phi)$ or $\sin(M\phi)$ dependence. Rotation and products of the real spherical harmonics lead then to real Wigner-D functions $D^L_{MK}$ and new Clebsch-Gordan coefficients denoted by $R^{L_3M_3}_{L_1M_1\,L_2M_2}$. The corresponding definitions are discussed in detail in Appendix~\ref{app:rr}.

\section{Hamiltonian}\label{sec:ham}

The Hamiltonian describing the full multi-mode Jahn-Teller problem with Coulomb interaction of the \csixtynminus {} ion has four terms: 
\begin{equation}\label{eq:H}
  H = H_p + H_{ep} + H_J + H_U
\end{equation} 
The first term is the energy of the 8 $d$-phonon multiplets
\begin{equation}
  H_p = \sum_{k\alpha} \omega_\alpha \left(a^\dagger_{k\alpha}a_{k\alpha} + \frac{1}{2}\right),
\end{equation} 
where $\omega_\alpha$ are the frequencies of the phonon multiplets~\cite{art:gunnarsson95b} ($\alpha=1\ldots 8$). $a^\dagger_{k\alpha}$ and $a_{k\alpha}$ are the phonon construction and annihilation operators. According to the definition in appendix~\ref{app:rr}, the quantum numbers $k=-2\ldots 2$ correspond to the $d$-symmetries $\sqrt 3 xy$, $\sqrt 3 yz$, $z^2-(x^2+y^2)/2$, $\sqrt 3 xz$, $\sqrt 3(x^2-y^2)/2$ respectively. As discussed in the introduction, we only consider linear electron-phonon coupling which is given by
\begin{equation}\label{eq:Hep}
  H_{ep} = -\sqrt {\frac{3}{2}} \sum_{\alpha\,k\,n\,n'\,s}
    \omega_\alpha g_\alpha\, R^{2k}_{1n\,1n'}\,
    c^\dagger_{ns}  c_{n's}\, \hat q_{k \alpha},
\end{equation} 
where $g_\alpha$ are the coupling constants and $\hat q_{k \alpha} = (a^\dagger_{k\alpha} + a_{k\alpha})/\sqrt 2$ the operators for the phonon coordinates. $c^\dagger_{ns}$ and $c_{n's}$ are the electron operators. They have spin $s$ and quantum numbers $n=-1,0,1$ which correspond to the $p$-symmetries $y,z,x$ respectively. $R^{2k}_{1n\,1n'}$ are the Clebsch-Gordan coefficients for the real spherical harmonics (see appendix~\ref{app:rr}). The resulting matrix elements of $H_{ep}$ are given in~(\ref{eq:Delta}). Note that both, the electron operators $c^\dagger_{ns}$ and $c_{n's}$ as well as the phonon operators $a^\dagger_{k\alpha}$, $a_{k\alpha}$ and $\hat q_{k \alpha}$ are tensor operators with rank 1 and 2 respectively. The sum in~(\ref{eq:Hep}) involving the Clebsch-Gordan coefficients is the simplest non-trivial scalar which can be built from rank 1 and rank 2 tensors. 

The last two terms in~(\ref{eq:H}) describe the Coulomb interaction among the $p$-electrons. The Coulomb interaction splits the charge states $N=2,3,4$ into multiplets characterized by the total electron angular momentum and the spin because spin-orbit coupling is omitted. All multiplet energies can be expressed by two parameters $J$ and $U$ giving rise to the two terms $H_J$ and $H_U$~\cite{art:wierzbowska04}. They have the form
\begin{eqnarray}
  H_J &=&
   \frac{J}{2}\sum_{nmss'}\left(c_{ns}^\dagger c_{ns'}^\dagger c_{ms'} c_{ms}+
    c_{ns}^\dagger c_{ms'}^\dagger c_{ns'}c_{ms}\right),\label{eq:HJ} \\
  H_U&=&\frac{U}{2}\,N(N-1), \label{eq:Hu}
\end{eqnarray}
where $N$ is the total number of electrons. In the literature, $H_J$ is referred to as the Hund's rule coupling and leads to the multiplet splittings listed in Tab.~\ref{tab:hund}. The second parameter,  $U$, is the overall charging energy. It is of the order $U=1-3$~eV, depending on the screening~\cite{art:wierzbowska04,art:pederson92}. $J$ is at least one order of magnitude smaller~\cite{art:wierzbowska04} as will be discussed in section~\ref{sec:c60spec}. In the following we will drop $H_U$ because we always work with a fixed number of electrons. 
\begin{table}
\begin{center}
\begin{tabular}{cccc|cccc}
  $N$ & $S$ & $L_e$ & $E\tl{mult}$ & $N$ & $S$ & $L_e$ & $E\tl{mult}$  \\
  \hline
  2,4 & 0 & 0 & $4J$ & 3 & 1/2 & 1 & 2J  \\
  2,4 & 0 & 2 & $J$  & 3 & 1/2 & 2 & 0   \\
  2,4 & 1 & 1 & $-J$ & 3 & 3/2 & 0 & -3J
 \end{tabular}
\caption{\label{tab:hund}
   Multiplet energies $E\tl{mult}$ of \csixtynminus{} for $N=2,3,4$ electrons which arise from the Hund's rule coupling $H_J$ given in~(\ref{eq:HJ}). The quantum numbers $N,S,L_e$ denote the number of electrons, the total spin and total electron angular momentum.
    }
\end{center}
\end{table}

All terms in the Hamiltonian~(\ref{eq:H}) are scalars and therefore SO(3) invariant. Hence, the total angular momentum $\V L=\V L_e+\sum_{\alpha}\V L_{p\alpha}$, which is the sum of the electron and phonon angular momenta, is conserved and eigenstates of $H$ have quantum numbers $L$ and $M$. In addition, $H$ is particle-hole symmetric if $H_U$ is neglected. Therefore it is enough to study the cases $N=1,2,3$.

The Hamiltonian above can be simplified by introducing one effective $d$-phonon multiplet instead of 8 multiplets. This results in the effective mode approximation (EMA) which was in detail investigated by O'Brien~\cite{art:obrien72}. The phonon operators for the effective mode are a superposition of the original modes:
\begin{equation}
  a_k = \sum_{\alpha=1}^8 u_\alpha a_{k \alpha}\quad \textrm{with}\quad
  \sum_{\alpha=1}^8  u_\alpha^2 =1.
\end{equation}  
The real coefficients $u_\alpha$ can be determined by the variational principle: If $\ket \Phi$ is a state which only contains excitations of the effective mode, then the following general relation holds (for a derivation see Ref.~\onlinecite{mythesis}):
\begin{equation}
    \min_{\{u_\alpha\}}\langle H\rangle_\Phi=
     \omega_0 -\frac{5}{2}\bar\omega + \bar\omega
     \langle H_p\tu{eff}+H_{ep}\tu{eff}\rangle_\Phi +
      \langle H_J+H_U\rangle_\Phi, 
\end{equation}
where $\omega_0=5/2\sum_\alpha \omega_\alpha$ is the total zero point energy and $\bar\omega$ the effective frequency given below. $\langle \cdot \rangle_\Phi$ denotes the expectation value with respect to the state $\ket\Phi$. $H_p\tu{eff}$ and $H_{ep}\tu{eff}$ are the energy and electron-phonon coupling for the effective multiplet:
\begin{eqnarray}
  H_p\tu{eff} &=& \frac{5}{2} +N_p,\quad\textrm{where}\quad 
  N_p = \sum_{k=-2}^2  a^\dagger_{k}a_{k}, \\
  H_{ep}\tu{eff} &=& -g\sqrt {\frac{3}{2}} \sum_{k\,n\,n'\,s}
    R^{2k}_{1n\,1n'}\, c^\dagger_{ns}  c_{n's}\, \hat q_{k},
\end{eqnarray} 
where $N_p$ is the phonon number operator for the effective mode and 
\begin{equation}\label{eq:effparams}
  g^2=\sum_{\alpha=1}^8 g_\alpha^2, \quad
  u_\alpha=\frac{g_\alpha}{g},\quad
  \bar \omega = \sum_{\alpha=1}^8  \omega_\alpha \, u_\alpha^2.
\end{equation} 
Hence, the ground state energy for the effective single
mode model with frequency $\bar\omega$ and coupling constant $g$
yields a variational estimate for the ground state
energy of the multi-mode problem. Parameters $\omega_\alpha$ and $g_\alpha$ are given in Tab.~\ref{tab:epparameters} and were taken from
Manini~\cite{thesis:Manini}. They go back to
photoemission measurements on gas-phase \csixtyminus{} by
Gunnarsson et al.~\cite{art:gunnarsson95,art:gunnarsson95b}. These parameters lead to $\bar\omega=72.1$~meV and $g=1.532$.
\begin{table}
\begin{center}
\begin{tabular}{l|c|c|c|c}
  Mode & $\omega_\alpha$ (cm$^{-1}$)
       & $\omega_\alpha$ (meV) & $\lambda_\alpha/N(0)$ (meV)
     & $g_\alpha$ \\
  \hline
  $H_g(8)$ & 1575 &  195.3 & 22 & .368 \\
  $H_g(7)$ & 1426 &  176.8 & 20 & .368 \\
  $H_g(6)$ & 1248 &  154.7 & 0  & .000 \\
  $H_g(5)$ & 1099 &  136.3 & 12 & .325 \\
  $H_g(4)$ & 772.5 &  95.8 & 16 & .448 \\
  $H_g(3)$ & 708.5 &  87.8 & 12 & .405 \\
  $H_g(2)$ & 430.5 &  53.4 & 38 & .924 \\
  $H_g(1)$ &  270 & 33.5   & 21 & .868 \\
  \hline
  Eff & 581 & 72.1 & -& 1.532
 \end{tabular}
\caption{\label{tab:epparameters}
  Frequencies and coupling constants for the vibrational modes in \csixty{} as
  taken from Manini~\cite{thesis:Manini}. The set of parameters originates from
  Gunnarsson~\cite{art:gunnarsson95b}. The parameter sets used
  in Ref.~\onlinecite{thesis:Manini} and~\onlinecite{art:gunnarsson95b} differ marginally in the
  frequencies. The coupling strength $g_\alpha$ and the
  electron-phonon coupling $\lambda_\alpha/N(0)$ are related by
  $g_\alpha^2=(6/5)\lambda_\alpha/(\omega_\alpha N(0))$. The last line are the resulting parameters for the effective mode approximation as given by equation~(\ref{eq:effparams}).  }
\end{center}
\end{table}

In the present work, we will always use the effective mode approximation. Hence, we will work with the Hamiltonian
\begin{equation} \label{eq:Heff}
  H\tl{eff}=H_p\tu{eff}+H_{ep}\tu{eff}+\frac{1}{\bar\omega}H_J,
\end{equation}  
where we dropped $H_U$ which only contributes a constant for fixed charge. For convenience, all energies are expressed in terms of $\bar\omega$. In order to shorten the notation, we will omit the superscript ``eff'' for $H_p$ and $H_{ep}$ in the following. The effective Hamiltonian has still SO(3) and particle-hole symmetry. The use of the EMA is justified for two reasons: First, it was shown by O'Brien that it is a good approximation for the ground state energy and that multimode corrections are small~\cite{art:obrien96}. This will also be confirmed in this work when comparing the present results to the literature in section~\ref{sec:c60spec}. Second, we are interested in the low energy excitations of the \csixtynminus{} ion. Generally, there are two types of excitations: rotator excitations involving a collective distortion corresponding to the effective mode and vibrational excitations involving ``individual'' modes out of the phonon spectrum. While the EMA is well suited for rotator excitations, it obviously doesn't capture vibrational excitations. While the former possess the energy scale $\bar\omega/(3g^2)\approx 10$~meV (see below), the vibrational excitations lie in the range $\omega_\alpha\approx 30-200$~meV. 
Consequently, the low-energy excitations are rotator excitations and are well described within the EMA. 

In what follows, it will be convenient to express the electron-phonon coupling term $H_{ep}=g(A^\dagger_{ep}+A_{ep})/\sqrt{2}$ in terms of the operator $A_{ep}$ defined by
\begin{equation}\label{eq:Aep}
  A_{ep} = -\sqrt{\frac{3}{2}} \sum_{k\,n\,n'\,s}
    R^{2k}_{1n\,1n'}\,
    c^\dagger_{ns}  c_{n's}\,  a_k.
\end{equation}
The operators $A_{ep}$ and $A_{ep}^\dagger$ can be understood as annihilation and
creation operators because they annihilate or create a phonon.
However, they don't obey simple commutation relations as $a^\dagger_k$ and $a_k$. For this reason, the $p\otimes d$ Jahn-Teller problem doesn't
have a simple analytic solution such as a single displaced harmonic oscillator.

\section{Coherent states}\label{sec:coh}

%
\begin{figure}
\begin{center}
\includegraphics[height=0.125\textheight]{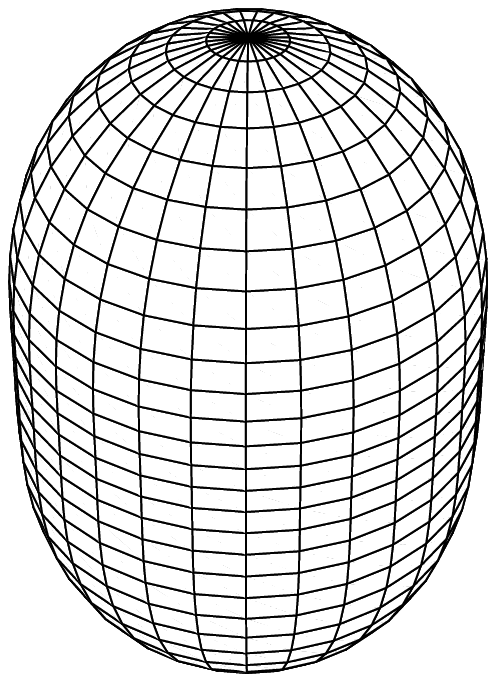}
\includegraphics[height=0.12\textheight]{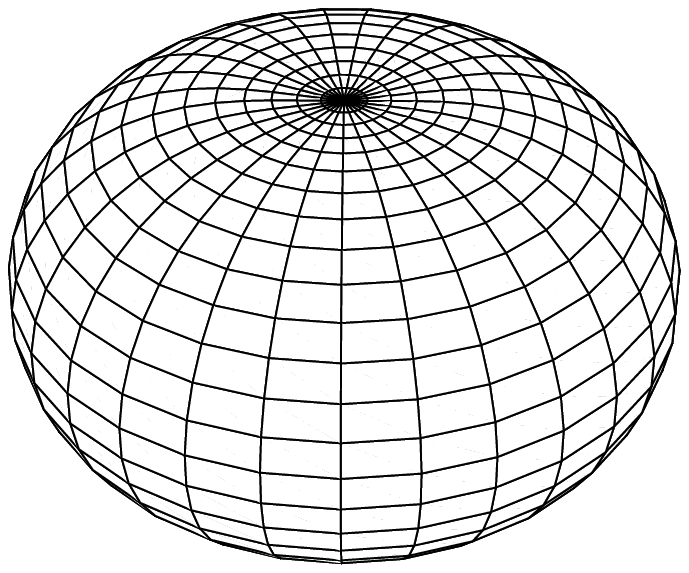}
\includegraphics[height=0.12\textheight]{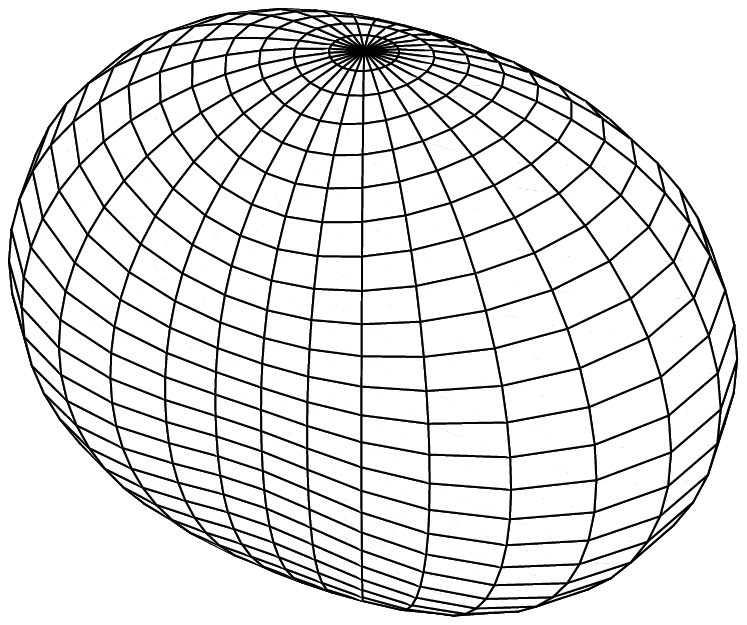}
\caption{\label{fig:distortedC60}
   Distortion of a sphere with radial displacement 
   $\Delta r(\Omega)\propto\cos\alpha\, Y_{20}(\Omega)+\sin\alpha\, Y_{22}(\Omega)$
   for shape parameters $\alpha=0,\pi,\pi/2$. The distortion $\Delta r$ 
   has symmetry $z^2-(x^2+y^2)/2$ for $\alpha=0,\pi$ and $\sqrt 3(x^2-y^2)/2$ for $\alpha=\pi/2$. The distorted spheres for $\alpha=0,\pi$ are axially symmetric with respect to $z$-axis. The case $\alpha=\pi/2$ is not axially symmetric, but has twofold axes given by the coordinate axes. }
\end{center}
\end{figure}
The Jahn-Teller problem under consideration involves distortions
of the molecule which we describe by coherent phonon states of the
type
\begin{equation}\label{eq:coh}
  \ket{\V q} =\exp\left(-i\sum_{k=-2}^2 q_k\hat p_k\right)\ket 0,
\end{equation}
where $\ket 0$ denotes the vacuum state and
$\hat p_k=i(a_k^\dagger-a_k)/\sqrt 2$ the phonon momentum operator. 
The vector $\V q=(q_{-2},\ldots,q_2)$ gives the
displacements of the oscillators and parameterizes the state. Generally, any 5-vector $\V q$ can be parameterized by the overall magnitude $q=|\V q|$, a shape parameter $\alpha$ and three Euler angles $\Theta=(\phi,\theta,\gamma)$~\cite{art:obrien96}:
\begin{equation}\label{eq:qparam}
  \V q=q \,D^2(\Theta)\,\V (0,0,\cos\alpha,0,\sin\alpha),
\end{equation}
where $D^2(\Theta)$ denotes a $5\times 5$ matrix with elements given by the Wigner D-functions $D^2_{kk'}(\Theta)$. 
The effect of the shape parameter $\alpha$ is illustrated in
Fig.~\ref{fig:distortedC60}.
The state $\ket{\V q}$ has the properties:
\begin{eqnarray}
 U(\Theta)\ket{\V q} &=& \ket{D^2(\Theta)\V q},\nonumber  \\ 
  a_k\ket{\V q}  &=& \frac{q_k}{\sqrt 2}\ket{\V q},\nonumber \\
  \bra{\V q}\hat q_k \ket{\V q} &=& q_k,\label{eq:cohprops} \\
  \langle \V q'\ket{\V q} &=& e^{-\frac{1}{4}|\V q-\V q'|^2}\nonumber \\
  \bra{\V q} H_{ep} \ket{\V q}_p &=& \sum_{nn's}
    \Delta_{nn'}(\V q)\, c^\dagger_{ns}c_{n's}. \nonumber
\end{eqnarray}
The first property involves the rotation operator $U(\Theta)$ and defines how the coherent state transforms under rotations. In the last equation, the expectation value of the electron-phonon coupling is 
determined with respect to the phonon state $\ket{\V q}$ which leaves an electron operator with matrix elements $\Delta_{nn'}(\V q)$ given by
\begin{equation}\label{eq:Delta}
  \Delta(\V q)=\frac{g}{2}\left(  \begin{array}{ccc}
     q_0+\sqrt 3 q_2 &-\sqrt 3 q_{-1}  &-\sqrt 3 q_{-2}  \\
      -\sqrt 3 q_{-1} & -2q_0 & -\sqrt 3 q_1 \\
     -\sqrt 3 q_{-2}  &-\sqrt 3 q_1 & q_0-\sqrt 3 q_2
  \end{array}\right) 
\end{equation} 
In order to study the eigenvalues of $\Delta(\V q)$, we invoke the SO(3) symmetry of the Hamiltonian. Due to this symmetry, the eigenvalues of $\Delta(\V q)$ and $\Delta(\V q')$ have to be equal if $\ket{\V q}$ and 
$\ket{\V q'}=U(\Theta)\ket{\V q}$ are related by a rotation. Therefore, in view of the parameterization~(\ref{eq:qparam}), the eigenvalues of $\Delta(\V q)$ only depend on the magnitude $q$ and the shape parameter $\alpha$. Choosing $\V q=(0,0,q\cos\alpha,0,q\sin\alpha)$ makes $\Delta(\V q)$ diagonal and yields the eigenvalues
\begin{equation}\label{eq:Deltaev}
  (\Delta_y,\Delta_z,\Delta_x)=gq\left(
     \cos\left[\alpha-\frac{\pi}{3}\right], 
     -\cos\alpha ,
     \cos\left[\alpha+\frac{\pi}{3}\right]\right)
\end{equation} 
%
%
As discussed below, the expectation value of the electron-phonon coupling term $H_{ep}$, i.e. the eigenvalues $\Delta_y$, $\Delta_z$, $\Delta_x$, determine the leading term of the Jahn-Teller energy gain. 

In contrast, the low-energy excitations emerging from the rotator physics have a much smaller energy scale which is given by the moment of inertia of state $\ket{\V q}$. Note that the moment of inertia in question is the one carried by the $H_g$ phonons and can be thought of as the moment of inertia carried by a tidal wave. It should not be confused with the moment of inertia for overall rotations of the molecule which does not enter the present problem. The moment of inertia can be obtained using a semi-classical approach~\cite{art:auerbach94,art:obrien96}. It can be shown that the semi-classical equations of motion for the five phonon coordinates coupled to the ionic charge take the form of a quantum rotator in the limit $g\to\infty$. This yields the moments of inertia. In the case $\V q=0$ where no phonons are excited, one has $\V L\ket{\V q}=0$ and therefore no moment of inertia. For $\V q\neq0$ the state $\ket{\V q}$ acquires moments of inertia which, due to the overall SO(3) symmetry of the Hamiltonian, depend only on the magnitude $q$ and the shape parameter $\alpha$ of the distortion $\V q$. For $\V q=(0,0,q\cos\alpha,0,q\sin\alpha)$ one has~\cite{art:auerbach94}:
\begin{equation}\label{eq:inertia}
  (I_y,I_z,I_x)=4q^2\left( 
  \cos^2\left[\alpha+\frac{\pi}{6}\right],\sin^2\alpha,
  \cos^2\left[\alpha-\frac{\pi}{6}\right]\right)
\end{equation} 
%

\section{\csixtyminus}\label{sec:c60m1}

The \csixtyminus{} ion is the simplest case among the \csixtynminus{} ions. In particular, the Hund's rule coupling term $H_J$ defined in~(\ref{eq:HJ}) is strictly zero and one only has to deal with the Jahn-Teller effect. In the following we attempt to construct variational wavefunctions starting from coherent states defined in~(\ref{eq:coh}). The idea is to choose the electron wavefunction and the distortion such as to minimize the electron-phonon coupling $H_{ep}$. This yields a state which gives the leading term of the Jahn-Teller energy gain. In a second step, the state is projected onto angular momentum subspaces in order to investigate the rotator excitations. Finally, we improve the variational estimate by enlarging the Hilbert space of the trial function.

Given the eigenvalues~(\ref{eq:Deltaev}), the electron-phonon coupling is minimized by putting the electron in the $z$-orbital ($n=0$) and choosing $\alpha=0$. In this case the electronic levels split into a singlet with energy $-gq$ and a doublet with energy $gq/2$ (neglecting spin degeneracy). This yields the trial wavefunction
\begin{equation}
  \ket{\Psi_0(q)}=c^\dagger_{0\uparrow}\ket{q\V e_0},
\end{equation}
where $\V e_0=(0,0,1,0,0)$ is a unit vector in the 5-dimensional phonon coordinate space. This state is normalized and an eigenstate of $A_{ep}$ with eigenvalue $-q/\sqrt 2$. The expectation value of $H$ is
\begin{equation}\label{eq:Psi0HPsi0}
  \bra{\Psi_0}H\ket{\Psi_0}=
  \frac{5}{2}+\frac{q^2}{2}-gq,
\end{equation}
which is minimal for $q=g$ and which yields an
upper bound for the ground state energy:
\begin{equation}\label{eq:vs.m1.sgse}
  E_{0}^{N=1}=\frac{5}{2}-\frac{g^2}{2}.
\end{equation}
The energy $-g^2/2$ is the leading term of the Jahn-Teller energy gain of the \csixtyminus{} ion. 

The variational estimate of the ground state energy above can be improved by enlarging the Hilbert space for the variational wavefunction. We consider three different Hilbert spaces spanned by the following choices of wavefunctions
\begin{eqnarray}
  (i)&&\ket{\Psi_0},\, H\ket{\Psi_0} \nonumber\\
  (ii)&&\ket{\Psi_0},\, A_{ep}^\dagger\ket{\Psi_0} \\
  (iii)&&\ket{\Psi_0},\, H\ket{\Psi_0},\, A_{ep}^\dagger\ket{\Psi_0} \nonumber
\end{eqnarray} 
Choice $(i)$ corresponds to a Lanzcos step. In choice $(ii)$, $A_{ep}^\dagger$ creates a phonon excitation. Choice $(iii)$ allows for both. The basis spanned by each choice depends on the variational parameter $q$ and the variational ground state energy is obtained upon minimization of the lowest eigenvalue with respect to $q$. Obviously, choice $(iii)$ must yield the lowest estimate. It turns out that choice $(ii)$ and $(iii)$ yield almost the same energies (rel. diff $<$0.2\%) whereas choice $(i)$ is somewhat higher (rel. diff to choice $(iii)$ $\sim$2\%). This is surprising as one would expect the Lanzcos choice to be optimal. The explanation is that energies are minimized with respect to $q$ and therefore the procedure is not a Lanzcos expansion in the proper sense. From these findings we deduce the following rule: Given a trial function $\ket 1$, we achieve a good improvement by adding state $\ket 2=A_{ep}^\dagger\ket 1$ to the Hilbert space of trial functions. Note that $A_{ep}^\dagger$ is a scalar and therefore $\ket 1$ and $\ket 2$ have the same symmetries. In addition, using $A_{ep}^\dagger$ instead $H$ to create a second state yields simpler wavefunctions. Below, we make extensive use of this rule. 

The state $\ket{\Psi_0}$ considered above is not an angular momentum state, but it can be understood as a rotator at rest. Its moments of inertia are given in~(\ref{eq:inertia}). For $\alpha=0$ and $q=g$ we find $I=I_x=I_y=3g^2$ and $I_z=0$. Therefore we expect a rotator spectrum given by $L(L+1)/(6g^2)$ which is on a smaller energy scale than the Jahn-Teller energy gain of the order $g^2$. The rotator takes up 2 degrees of freedom of the 5-dimensional phonon space. The other three degrees of freedom are vibrations of the rotator. This picture emerges also when treating the $p\otimes d$ Jahn-Teller problem semiclassically as done in Ref.~\onlinecite{art:auerbach94,art:obrien96}. These references show that the asymptotic rotator spectrum in the $g\to\infty$ limit becomes 
\begin{equation}\label{eq:n1asym}
  E_\infty^{N=1}=-\frac{g^2}{2} +\frac{3}{2} +\frac{L(L+1)}{6g^2}.
\end{equation} 
The first term is the leading Jahn-Teller energy gain.
The second term is the zero-point energy of the three remaining decoupled oscillators and the last term are the rotator excitations. 

Before going into the details of the projection technique, we need to investigate the symmetries of $\ket{\Psi_0}$ in order to know which projections are non-zero. 
First we note that $L_z\ket{\Psi_0}=0$ because $\ket{\Psi_0}$ is constructed from operators $c^\dagger_{0\uparrow}$ and $a^\dagger_0$ for which the $z$-component of angular momentum vanishes. Hence, only $M=0$ projections are non-zero. In addition, due to the electron operator $c^\dagger_{0\uparrow}$, $\ket{\Psi_0}$ is odd under a $\pi$-rotation around the $y$-axis. Therefore only projections with odd total angular momentum $L$ are allowed as can formally be shown using the second projection operator property given in~(\ref{eq:Qprops}). These findings agree with the literature~\cite{art:auerbach94,art:obrien96}. 

Projection operators onto angular momentum subspaces are given by
\begin{equation}\label{eq:QLMK}
  Q^L_{MK}=\frac{2L+1}{8\pi^2}\int d\Theta\,D^L_{MK}(\Theta)\, U(\Theta),
\end{equation} 
where $U(\Theta)$ is the rotation operator and $D^L_{MK}(\Theta)$ the real Wigner D-function. $\int d\Theta=\int_{0}^{2\pi}d\phi\int_{0}^{\pi}d\theta\sin\theta
\int_{0}^{2\pi}d\gamma$ denotes the integration over the Euler angles $(\phi,\theta,\gamma)$. The prefactor serves for proper normalization and arises from the orthogonality relation of the Wigner-D functions~\cite{book:QToAngularMomentum}. These projection operators have the following properties with respect to normalized angular momentum states $\ket{LM}$
\begin{equation}
    \bra{L_1M_1}Q^L_{MK}\ket{L_2M_2}=
    \delta_{LL_1}\,\delta_{LL_2}\,\delta_{MM_1}\,\delta_{KM_2}.
\end{equation} 
The lowest state is expected for $L=1$ which leads to the trial function
\begin{equation}\label{eq:P0}
  \ket{P_0(q )}=Q^1_{00}\,\ket{\Psi_0(q)}=
   Q^1_{00}\,c^\dagger_{0\uparrow}\ket{q\V e_0}.
\end{equation}
We have adopted the letter $P$, as in atomic physics, to
indicate that the state has total angular momentum $L=1$. The
spin degeneracy is not indicated because it is always 2 in the
case of one electron. Note that this trial function was already proposed in Ref.~\onlinecite{art:sookhun03}, but without having $q$ as a variational parameter. 

\begin{figure}
\begin{center}
\includegraphics[width=0.5\textwidth]{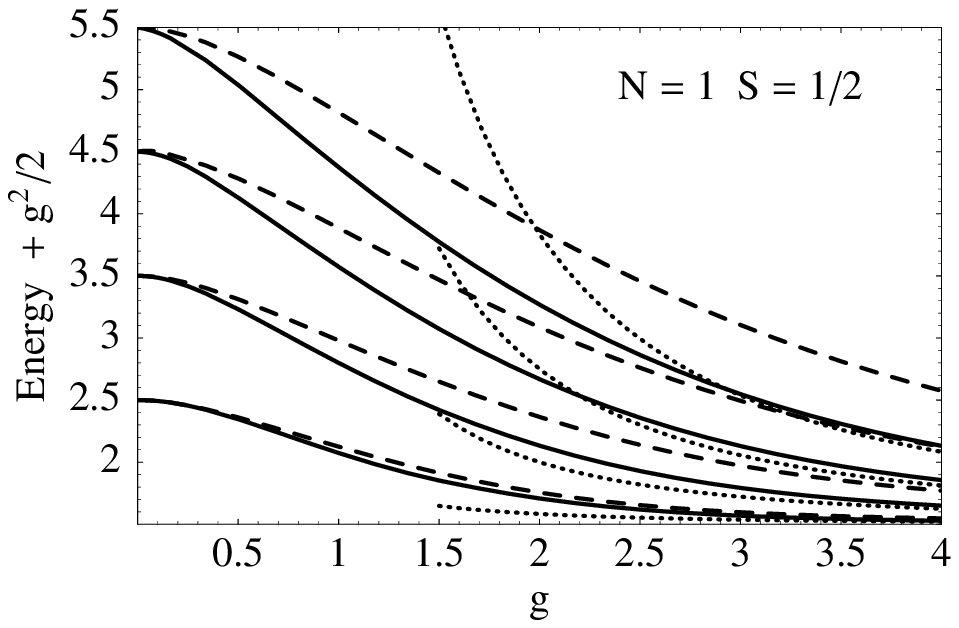}
\includegraphics[width=0.5\textwidth]{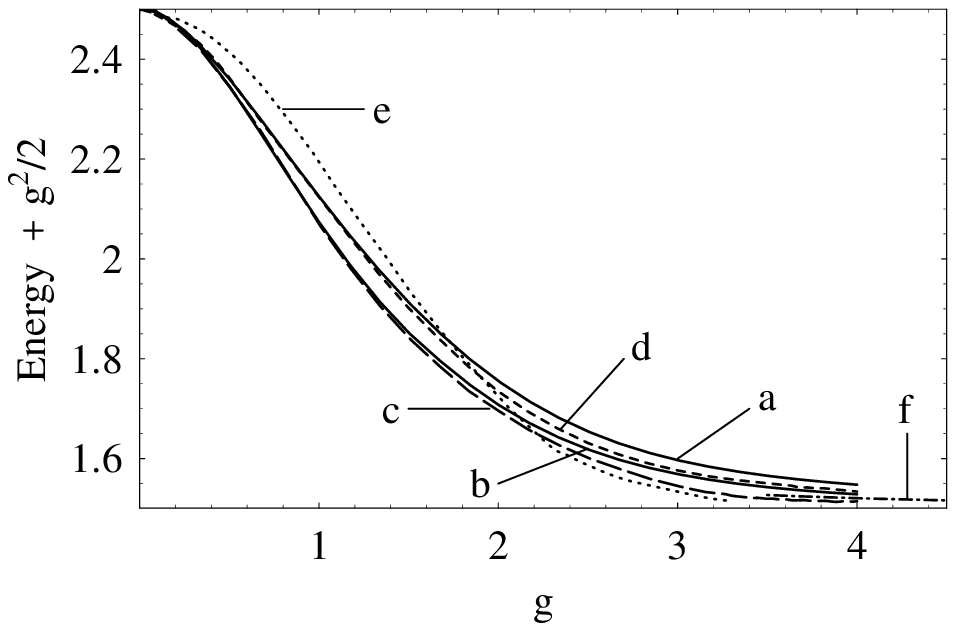}
\caption{\label{fig:c60minusnrg}
  \emph{Upper panel:} Ground state ($L=1$) energy and rotator excitations ($L=3,5,7$)
  for \csixtyminus{} as a function of the coupling strength $g$
  and relative to the asymptotic energy $-g^2/2$.
  The dashed lines are the variational
  energies for states as given
  by~(\ref{eq:P0}), but with an appropriate projector $Q^L_{00}$.
  The solid lines are the energies of the improved variational approach.
  The dotted lines are asymptotic behavior given
  in~(\ref{eq:n1asym}).
  \emph{Lower panel:}
  Energy of the ground state of \csixtyminus{}. 
  Solid lines (a,b): Present results, the lower curve (b) is the improved version. Long dash (c): Numerical result of O'Brien~\cite{art:obrien71}. Short dash (d): Variational wavefunction of Dunn et al. (curve (b) in Ref.~\onlinecite{art:dunn02}). Dots (e): Approximate analytical result of Chancey~\cite{art:chancey87}. Dash-dots (f): Asymptotic behavior given in~(\ref{eq:n1asym}). 
}
\end{center}
\end{figure}

Since $\ket{\Psi_0}$ is an eigenvector of $A_{ep}$, $\ket{P_0}$ is an eigenvector as well because $A_{ep}$ is a scalar and $[A_{ep},Q^1_{00}]=0$. Hence, we have  
$\bra{P_0} H_{ep} \ket{P_0}/\langle {P_0} \ket{P_0}= -gq$.
The calculation of the expectation value of the phonon number $N_p$ is more involved. The norm of the wavefunction $\ket{ P_0}$ is given by
\begin{eqnarray} \label{eq:NormP0}
  \langle {P_0} \ket{P_0}&=&
     \bra{\Psi_0}\,Q^1_{00}\, \ket{\Psi_0} \\ \nonumber
   &=&\frac{3}{8\pi^2}\int d\Theta\,D^1_{00}(\Theta)\,
      \bra{\Psi_0}\, U(\Theta)\, \ket{\Psi_0}, 
\end{eqnarray}
In the first equation we used the projection operator property $(Q^1_{00})^2=Q^1_{00}$. As discussed in the introduction, this step is crucial because eliminating a projection operator eliminates an integration over Euler angles. In the second equation of~(\ref{eq:NormP0}) the definition of $Q^1_{00}$ is substituted. The integral above involves the matrix element
\begin{equation} \label{eq:MENormP0}
    \bra{\Psi_0}\, U(\Theta)\, \ket{\Psi_0} =
    \cos\theta \exp\left(-\frac{3}{4}\,q^2\sin^2\theta\right)
\end{equation}
which can be calculated using the rotation rule for tensor operators~(\ref{eq:Trot}) as well as properties~(\ref{eq:cohprops}). 
The expectation value $\bra{P_0} N_p \ket{P_0}$ is evaluated likewise and involves the matrix element 
\begin{eqnarray}\label{eq:MEExpValNpP0}
  \lefteqn{\bra{\Psi_0}\,N_p\, U(\Theta)\, \ket{\Psi_0} =} \\
   &&\frac{q^2}{2} \cos\theta
     \left(\frac{3}{2}\,\cos^2\theta-\frac{1}{2}\right)
     \exp\left(-\frac{3}{4}\,q^2\sin^2\theta\right), \nonumber
\end{eqnarray}
where it should be noted that $[N_p, U(\Theta)]=0$ because $N_p$ is a scalar. The matrix elements (\ref{eq:MENormP0}) and (\ref{eq:MEExpValNpP0}) don't depend on the Euler angles $\phi$ and $\gamma$ which leaves one integration over $\theta$. Substituting $t=\cos\theta$ we find  
\begin{equation}\label{eq:defh}
  \frac{\bra{P_0} N_p \ket{P_0}}{\langle {P_0} \ket{P_0}}
   = \frac{q^2}{2}
    \frac{\int_{-1}^1 t^2\left(\frac{3}{2}t^2- \frac{1}{2}\right)e^{-\frac{3}{4}q^2(1-t^2)}}
         {\int_{-1}^1 t^2e^{-\frac{3}{4}q^2(1-t^2)}}
   \equiv  \frac{q^2}{2}h(q).
\end{equation}
Note that the integrals can be expressed in terms of error functions.
The last equality defines the function $h(q)$ which varies
smoothly from $h(0)=2/5$ to $h(\infty)=1$. Putting the different parts together,
the expectation value of $H$ becomes
\begin{eqnarray} \label{eq:vs.c60m1.dd.Ed}
  E^{N=1}(g,q)=\frac{\bra{P_0} H \ket{P_0}}{\langle {P_0} \ket{P_0}} =
  \frac{5}{2} + \frac{q^2}{2}\,h(q) -g q.
\end{eqnarray}
The only difference to the energy found above
is the factor $h(q)$ which renormalizes the phonon energy $q^2/2$.
This additional energy gain is due to 
the ``delocalization'' of the distortion in the projected state. Minimizing $E_0(g,q)$  with respect to $q$ for a given coupling strength $g$ yields the variational ground state energy shown in Fig.~\ref{fig:c60minusnrg}. 

Higher rotator excitations are constructed by replacing the projector $Q^1_{00}$ in~(\ref{eq:P0}) with a projector $Q^L_{00}$ on a higher angular momentum space. This results in replacing the Wigner D-function $D^1_{00}$ in integral~(\ref{eq:NormP0}) with $D^L_{00}$. Note that the Wigner D-functions $D^L_{00}(\Theta)=P_L(\cos\theta)$ are given by the Legendre polynomials $P_L$. Results for $L=3,5,7$ are shown in Fig.~\ref{fig:c60minusnrg}. As can be seen, the asymptotic behavior for $g\to\infty$ differs substantially from the semiclassical result~(\ref{eq:n1asym}). In order to improve these energies we follow the rule suggested above. That is, we use a 2-dimensional Hilbert space spanned by $\ket{P_0}$ and $A_{ep}^\dagger\ket{P_0}$ and minimize~\cite{note:thesispaper1} the lower eigenvalue with respect $q$. Higher angular momentum excitations are treated likewise. Note that $A_{ep}^\dagger$ is a scalar and therefore $[Q^L_{MK},A_{ep}^\dagger]=0$. Results are shown in Fig.~\ref{fig:c60minusnrg}. The improvement is substantial and the asymptotic behavior for $g\to\infty$ fits well the semi-classical result~(\ref{eq:n1asym}).

Comparison of the present result for the ground state energy of \csixtyminus{} with other works  are shown in the lower panel of Fig.~\ref{fig:c60minusnrg}. Our improved result fits well the numerical calculation of O'Brien~\cite{art:obrien71} in the low coupling regime. It deviates for higher couplings where the energies of O'Brien are somewhat lower. This might be due to the fact that our approach is strictly variational whereas in Ref.~\onlinecite{art:obrien71} the matrix elements for states with a large number of phonons are extrapolated which may lead to non-variational energies. In the limit $g\to\infty$, our result approaches the asymptotic behavior~(\ref{eq:n1asym}) smoothly from above.   

\section{\csixtytwominus}

A new aspect of \csixtytwominus{} is the non-trivial Hund's rule coupling $H_J$ given in~(\ref{eq:HJ}). The main features of the competition between Jahn-Teller effect and Hund's rule coupling can be observed on the level of unprojected states which we discuss first. As before we start with the following state which minimizes the electron-phonon coupling $H_{ep}$:
\begin{equation}
    \ket{{}^1\Psi_0(q)}=c_{0\uparrow}^\dagger c_{0\downarrow}^\dagger
      \ket{q\V e_0}
\end{equation}
The upper index indicates that the state is a spin singlet with spin degeneracy 1. This state is normalized and an eigenvector of $A_{ep}$ with eigenvalue $\sqrt 2q$ which is twice bigger than for \csixtyminus{} due to presence of 2 electrons. The expectation value of $H_J$ with respect to $\ket{{}^1\Psi_0}$ is $2J$ which implies, in view of the multiplet energies given in Tab.~\ref{tab:hund}, that $\ket{{}^1\Psi_0}$ is not an eigenstate of $H_J$. In fact, $H_J$ couples $\ket{{}^1\Psi_0}$ to another state 
\begin{equation}
    \ket{{}^1\Psi_1(q)}=\frac{1}{\sqrt 2}
      \left(c_{1\uparrow}^\dagger c_{1\downarrow}^\dagger
       + c_{-1\uparrow}^\dagger
       c_{-1\downarrow}^\dagger \right)\ket{q\V e_0},
\end{equation}
which also has spin $S=0$ and total angular momentum $L_z\ket{{}^1\Psi_1}=0$. 
It is an eigenstate of $A_{ep}$ with positive eigenvalue $q/\sqrt 2$. The two states $\ket{{}^1\Psi_0}$ and $\ket{{}^1\Psi_1}$ form a basis in which $A_{ep}$ is diagonal but $H_J$ is not. Eigenstates of $H_J$ are obtained by the following orthogonal transformation:   
\begin{eqnarray} \label{eq:c60m2statbasis}
    \ket{{}^1\Psi_S}&=&
       \frac{\ket{{}^1\Psi_0}+\sqrt 2\ket{{}^1\Psi_1}}{\sqrt 3}=
       \frac{1}{\sqrt 3} \sum_{n=-1}^1
        c_{n\uparrow}^\dagger c_{n\downarrow}^\dagger\ket{q\V e_0},\nonumber \\
   \ket{{}^1\Psi_D}&=&
       \frac{-\sqrt 2\ket{{}^1\Psi_0}+\ket{{}^1\Psi_1}}{\sqrt 3}.
\end{eqnarray}
The corresponding eigenvalues are $4J$ and $J$ respectively, as given in Tab.~\ref{tab:hund}. These two states are denoted with the lower indices $S$ and $D$ because they are eigenvectors of the total electron angular
momenta $\V L_e^2$ with angular momentum $L_e=0$ and  $L_e=2$ respectively. The Hamiltonian in the basis $({}^1\Psi_S,{}^1\Psi_D)$ takes the form
\begin{eqnarray}\label{eq:c60m2Hred}
   \frac{5}{2} +
   \frac{q^2}{2}+
     \left(\begin{array}{cc}
      4J/\bar\omega& \sqrt 2 gq\\
      \sqrt 2 gq & -gq +J/\bar\omega
     \end{array}\right). 
\end{eqnarray}
Minimizing with respect to $q$ yields the ground state energy
\begin{equation}\label{eq:E0N2}
  E_0^{N=2}=\frac{5}{2}+\frac{J}{\bar\omega}
    -g^2\,f\left(\frac{J}{\bar\omega g^2}\right),
\end{equation}
where the function $f$ decreases monotonically from $f(0)=2$ to $f(\infty)=1/2$. Hence, in the absence of Hund's rule coupling ($J=0$), the Jahn-Teller energy gain is $2g^2$. This is 4 times bigger than in \csixtyminus{} because the electron-phonon coupling is doubled due to the presence of two electrons. In the case of dominating Hund's rule coupling, i.e. $J\gg \bar\omega g^2$, the Jahn-Teller energy gain is reduced to $g^2/2$ but not entirely suppressed because $\bra{{}^1\Psi_D}H_{ep}\ket{{}^1\Psi_D}=-gq$.  Note that this finding differs from the general belief that strong Hund's rule coupling completely suppresses the Jahn-Teller effect.

If we don't restrict our view on the $S=0$ sector, then, of course, the spin triplet state will be favored for large enough and positive $J$ (see Tab.~\ref{tab:hund}). Hence, there is a level crossing between low and high spin state which depends on the parameters $g$ and $J$. The Jahn-Teller problem in the $S=1$ sector is equivalent to the Jahn-Teller problem of \csixtyminus{} due to particle-hole symmetry. This is obvious when looking at the case of maximal spin $S_z=1$, where the spin-up states are occupied by two electrons and one hole. Therefore, the energy of the triplet state on this level of approximation is $E=5/2-g^2/2-J/\bar\omega$. This energy can be compared to the energy of the singlet state~(\ref{eq:E0N2}). One finds that the level crossing occurs at $J/(\bar\omega g^2)=0.5284$. This criterion is only little modified when going to projected variational states below (see Fig.~\ref{fig:c60xing}).

As before, we calculate rotator excitations using projection operators. The phonon coherent state in the present case is the same as for \csixtyminus{}. Therefore we expect the same rotator physics in the strong coupling limit $g\to\infty$. The moment of inertia of the rotator is $I=3q^2$. In the absence of Hund's rule coupling, where $q=2g$, we obtain $I=12g^2$ which implies that the energy scale of the rotator excitations in \csixtytwominus{} is four times smaller than in \csixtyminus{}. The full asymptotic behavior for $g\to\infty$ and $J=0$ is again obtained from the semiclassical approach~\cite{art:obrien96}: 
\begin{equation}\label{eq:n2asym}
  E_\infty^{n=2}\approx -2g^2+\frac{3}{2}+\frac{1}{12g^2}+\frac{L(L+1)}{24g^2}.
\end{equation}
As in \csixtyminus{}, the unprojected states are annihilated by $L_z$ which means that only $M=0$ projections are allowed. On the other hand, the states are invariant under $\pi$-rotations around the $y$-axis which requires $L$ to be even (see properties~(\ref{eq:Qprops}) of the projection operator). 

\begin{figure}
\begin{center}
\includegraphics[width=0.5\textwidth]{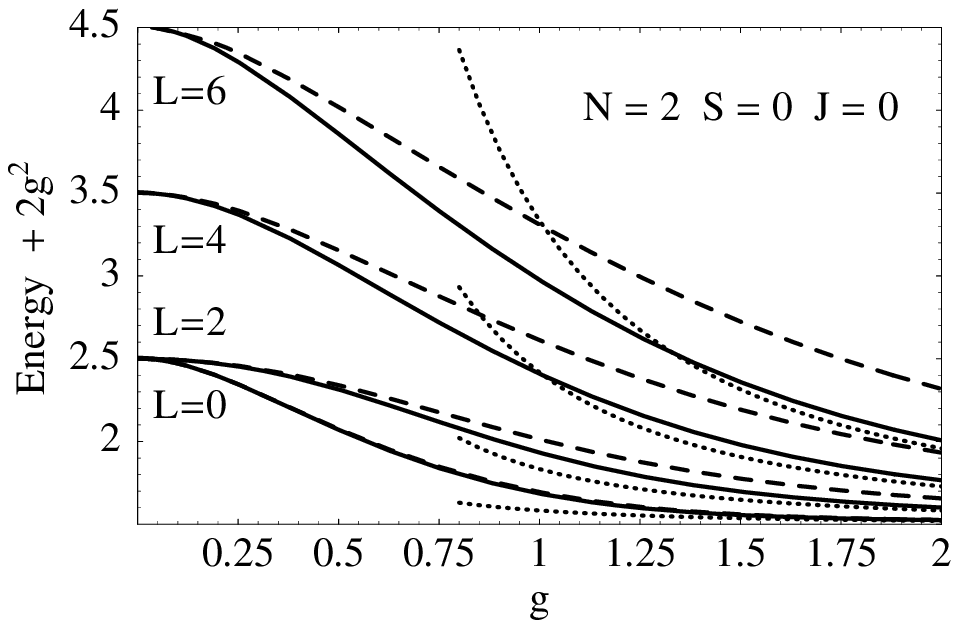}
\includegraphics[width=0.5\textwidth]{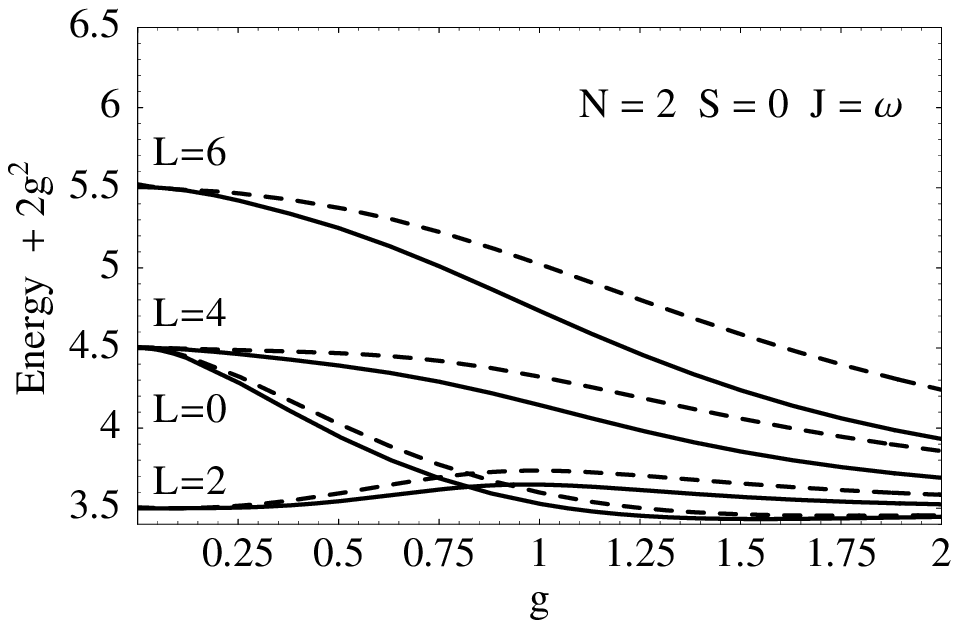}
\caption{\label{fig:c60m2nrg}
  \emph{Upper panel:} Rotator states $L=0,2,4,6$
  of \csixtytwominus{}
  in the singlet subspace ($S=0$) and for $J=0$. The levels are
  plotted as a function of the coupling strength $g$ and relative to the
  asymptotic energy $-2g^2$. Dashed lines are the results using basis~(\ref{eq:c60m2dynbasis}). Solid lines correspond to the
  improved approach and the dotted lines are the asymptotic behavior given in~(\ref{eq:n2asym}). 
  \emph{Lower panel:} As in the upper panel but with
  $J=\bar\omega$.
}
\end{center}
\end{figure}

In order to investigate the rotator physics in the presence of Hund's rule coupling we use the following projected states:
\begin{equation}\label{eq:c60m2dynbasis}
  \ket{{}^1X_S}=Q^L_{00}\ket{{}^1\Psi_S},\quad \ket{{}^1X_D}=Q^L_{00}\ket{{}^1\Psi_D},
\end{equation}   
where $X=S,D,G,I,\ldots$ stands for the letter denoting the total angular momenta $L=0,2,4,\ldots$ The two states $\ket{{}^1\Psi_S}$ and $\ket{{}^1\Psi_D}$ are eigenstates of $H_J$ with different eigenvalues and so are the two states defined in~(\ref{eq:c60m2dynbasis}) because $[H_J,Q^L_{00}]=0$. Therefore, they form an orthogonal basis. The calculation of the various matrix elements follows the procedure described above. Within basis~(\ref{eq:c60m2dynbasis}) and for a given $L$, the Hamiltonian has the matrix elements
%
\begin{eqnarray}
   \frac{\bra{{}^1X_S} H \ket{{}^1X_S}}{\langle {{}^1X_S} \ket{{}^1X_S}}&=&
   \frac{5}{2} +\frac{q^2}{2}\frac{F^L_1}{F^L_0}+\frac{4J}{\bar\omega}, \\
   \frac{\bra{{}^1X_D} H \ket{{}^1X_D}}{\langle {{}^1X_D} \ket{{}^1X_D}}&=&
   \frac{5}{2} +\frac{q^2}{2}\frac{F^L_2}{F^L_1}-gq+\frac{J}{\bar\omega},\nonumber\\
   \frac{\bra{{}^1X_D} H \ket{{}^1X_S}} 
      {\sqrt{\langle {{}^1X_S} \ket{{}^1X_S}\langle {{}^1X_D} \ket{{}^1X_D}}}&=&
      \frac{gq}{\sqrt 2}\left(\sqrt{\frac{F^L_1}{F^L_0}}+
	        \sqrt{\frac{F^L_0}{F^L_1}}\right), \nonumber
\end{eqnarray}
where the function $F^L_n(q)$ is defined in terms of Legendre Polynomials $P_L$: 
\begin{equation}
  F^L_n(q)=\frac{2L+1}{2}\int_{-1}^1 dt\, P_L(t)\,\left[P_2(t)\right]^n\,e^{-\frac{3}{4}q^2(1-t^2)}.
\end{equation} 
The energies which result from minimizing the lower eigenvalue are shown in Fig.~\ref{fig:c60m2nrg} for $J=0$ and $J=\bar\omega$. As expected, $J$ leads to an inversion of the $L=0$ and $L=2$ levels for small enough $g$. For $J=\bar\omega$ and $g=0$ the $L=0$ level has energy $4.5\bar\omega$. This state corresponds to an electronic $D$-state with one phonon excitation. The pure electronic $S$-state with no phonon excitation has higher energy $6.5\bar\omega$. For large enough $g$ the two spectra become very similar except for an overall energy shift $\sim 2J$. This corresponds to the expectation of $H_J$ with respect to state $\ket{{}^1\Psi_0}$ which minimizes the electron-phonon coupling.

As can be seen from Fig.~\ref{fig:c60m2nrg}, the asymptotic energies as calculated above overestimate the rotator excitations substantially. Therefore we improve the variational approach as discussed above. That is, the 2-dimensional Hilbert spanned by~(\ref{eq:c60m2dynbasis}) is enlarged to a 4-dimensional space using, in addition, the two states $A_{ep}^\dagger\ket{{}^1X_S}$ and $A_{ep}^\dagger\ket{{}^1X_D}$. The improved energies~\cite{note:thesispaper1} are also shown in Fig.~\ref{fig:c60m2nrg}. The asymptotic behavior agrees well with the semiclassical result~(\ref{eq:n2asym}).

For big enough Hund's rule coupling $J$, a level crossing occurs such that the $S=1$ spin triplet state becomes the ground state. The crossing between the spin singlet and spin triplet ground state defines a line in the $(g,J)$ parameter space as shown in Fig.~\ref{fig:c60xing}. Note that singlet-triplet crossing occurs always for smaller $J$ than the $L=0$ to $L=2$ crossing in the spin singlet sector. The energy of the triplet state is obtained by subtracting $J$ from the ground state energy of \csixtyminus{}. The crossing line was calculated using improved variational approaches. As can be seen in Fig.~\ref{fig:c60xing}, the criterion $J/(\bar\omega g^2)=0.5284$ derived above for the unprojected states becomes correct in the large $g$ limit. For $g\to 0$ the line ends at $J/(\bar\omega g^2)=3/4$ which can be shown using the perturbative results~\cite{art:auerbach94} for small $g$. 

\section{\csixtythreeminus}\label{sec:c60m3}

In order to find a state which minimizes $H_{ep}$ we start with the distortion $\V q=q(0,0,\cos\alpha,0,\sin\alpha)$ and put two electrons in the $x$-orbital and one in the $z$-orbital. According to~(\ref{eq:Deltaev}), the expectation value of $H_{ep}$ is $2\Delta_x+\Delta_z=-\sqrt 3 gq\sin\alpha$ which is minimal for $\alpha=\pi/2$. This yields the spin-1/2 state
\begin{equation} \label{eq:2Psi0}
  \ket{{}^2\Psi_0(q)}=c_{0\uparrow}^\dagger 
    c_{1\uparrow}^\dagger c_{1\downarrow}^\dagger \ket{q\V e_2},
\end{equation}  
where the upper index denotes the spin degeneracy and $\V e_2=(0,0,0,0,1)$. State $\ket{{}^2\Psi_0}$ is an eigenstate of $A_{ep}$ with eigenvalue $-\sqrt{3/2}q$. However, it is not an eigenstate of $H_J$.  $H_J$ couples $\ket{{}^2\Psi_0}$ to another state
\begin{equation} \label{eq:2Psi1}
  \ket{{}^2\Psi_1(q)}=c_{0\uparrow}^\dagger 
    c_{-1\uparrow}^\dagger c_{-1\downarrow}^\dagger \ket{q\V e_2},
\end{equation}  
which is an eigenstate of $A_{ep}$ with positive eigenvalue $\sqrt{3/2}q$. Eigenstates of $H_J$ are given by
\begin{eqnarray} \label{eq:c60m3statbasis}
    \ket{{}^2\Psi_P}=
       \frac{\ket{{}^2\Psi_0}+\ket{{}^2\Psi_1}}{\sqrt 2}, \quad
   \ket{{}^2\Psi_D}=
       \frac{-\ket{{}^2\Psi_0}+\ket{{}^2\Psi_1}}{\sqrt 2}.
\end{eqnarray}
These states are also eigenstates of the total electron angular momentum as indicated by the lower indices $P$ and $D$. Within basis~(\ref{eq:c60m3statbasis}), the Hamiltonian has the form
\begin{eqnarray}\label{eq:c60m3Hred}
   \frac{5}{2} +
   \frac{q^2}{2}+
     \left(\begin{array}{cc}
      2J/\bar\omega& \sqrt 3 gq\\  
      \sqrt 3 gq & 0
     \end{array}\right). 
\end{eqnarray}
Minimizing the lower eigenvalue with respect to $q$ yields
\begin{equation}\label{eq:c60m3nrgred}
  E_0^{n=3}=\left\{\begin{array}{ll} \displaystyle 
 	\frac{5}{2} & J> 3\bar\omega g^2, \\  & \\
	\displaystyle 
	\frac{5}{2}-\frac{3g^2}{2}+\frac{J}{\bar\omega}-\frac{J^2}{6\bar\omega^2 g^2}
	   & 0<J\leq 3\bar\omega g^2.
             \end{array}
  \right. 
\end{equation} 
For $J=0$ the Jahn-Teller energy gain is $-3g^2/2$ and somewhat reduced compared to \csixtytwominus{}. For $J> 3\bar\omega g^2$ the Jahn-Teller effect is completely suppressed. 

As in \csixtytwominus{}, there is a high spin  state ($S=3/2$) which is favored by the Hund's rule coupling. In this state, the electrons have parallel spin. Hence, each $p$-orbital is occupied by one electron and no Jahn-Teller coupling is possible. According to Tab.~\ref{tab:hund}, the ground state energy of the $S=3/2$ state is simply given by $5/2-3J$. Using energy~(\ref{eq:c60m3nrgred}) one finds the criterion 
$J/(\bar\omega g^2)=3(4-\sqrt{15})=0.381$ for the low-spin/high-spin  crossing. 

\begin{figure}
\begin{center}
\includegraphics[width=0.5\textwidth]{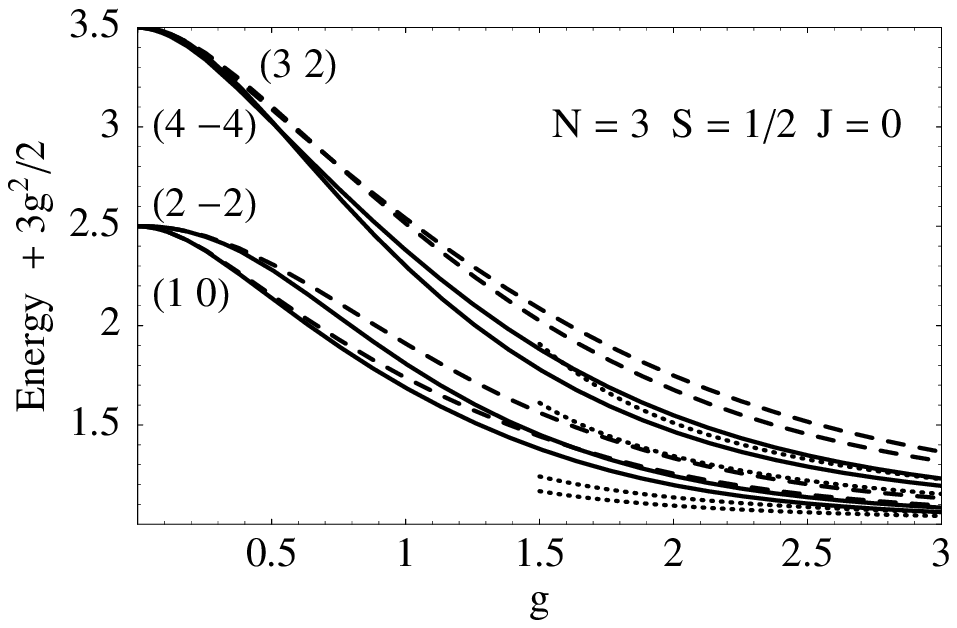}
\includegraphics[width=0.5\textwidth]{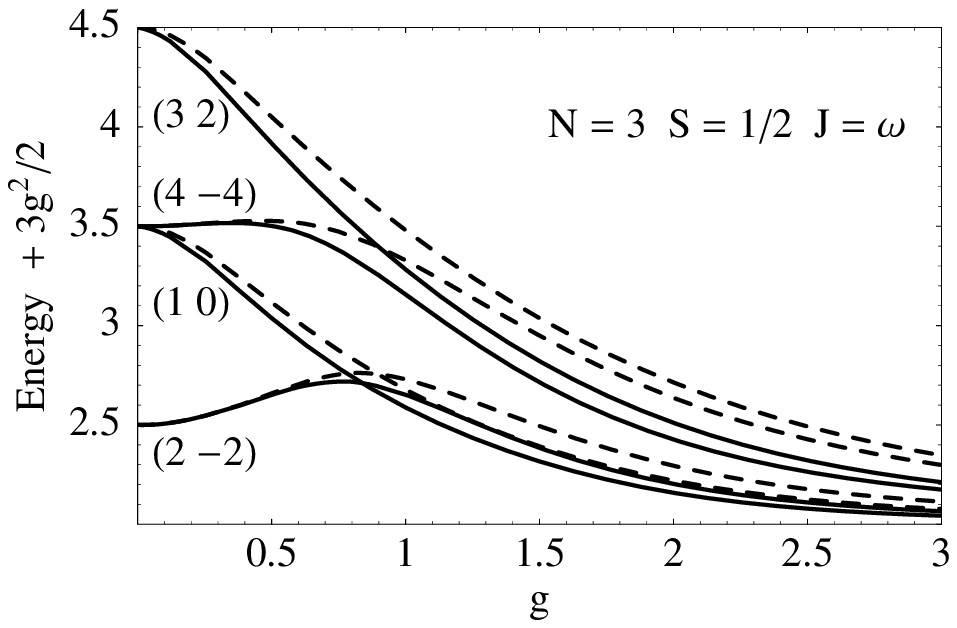}
\caption{\label{fig:c60m3nrg}
  \emph{Upper panel:} Rotator states $(L\,K)=(1\,0),(2\,-2),(4\,-4),(3\,2)$ of \csixtythreeminus{}
  in the $S=1/2$ sector and for $J=0$. The levels are
  plotted as a function of coupling strength $g$ and relative to the
  asymptotic energy $-3g^2/2$. Dashed lines are the result using basis~(\ref{eq:c60m3dynbasis}). Solid lines correspond to the
  improved approach and the dotted lines are the asymptotic behavior given in~(\ref{eq:n3asym}). 
  \emph{Lower panel:} As in the upper panel but with
  $J=\bar\omega$.
}
\end{center}
\end{figure}

In order to proceed with the projection, we first analyze the moments of inertia. According to relation~(\ref{eq:inertia}), there are two different moments of inertia $I_1=I_x=I_y=q^2$ and $I_3=I_z=4q^2$. The fact that $I_x=I_y$ and $I_z\neq0$ suggests that the rotational degrees of freedom behave similar to a symmetric top~\cite{art:auerbach94,art:obrien96}. The solution of the quantum mechanical symmetric top is well known (for an early review see Ref.~\onlinecite{art:dennison31}). The symmetric top has three rotational degrees of freedom, the Euler angles, and three conserved quantities: the rotational frequency around the principal axis $I_3$ of the top (quantum number $K$) and the angular momentum corresponding to the precession of this axis (quantum numbers $L,M$). The two remaining degrees of freedom become decoupled oscillators, i.e. motions of the top axis.  
The spectrum of the symmetric top~\cite{art:dennison31} is given by $(I_1^{-1}L(L\!+\!1)+[I_3^{-1}\!-\!I_1^{-1}]K^2)/2$. This spectrum is recovered for the low energy excitations of \csixtythreeminus{} in the limit $g\to\infty$ where $q=\sqrt 3 g$, $I_1=3g^2$ and $I_3=12g^2$. Within the semiclassical approach~\cite{art:obrien96} one finds the asymptotic behavior
\begin{equation}\label{eq:n3asym}
  E_{\infty}^{n=3}= -\frac{3g^2}{2}+1+\frac{1}{24g^2}
    +\frac{L(L+1)-\frac{3}{4}K^2}{6g^2}.
\end{equation}
Eigenfunctions of the top are given by the Wigner-D functions $D^L_{MK}$. Since we chose the distortion $\V q$ such that the $I_3$-axis corresponds to the $z$-axis, states with good quantum numbers $L,M,K$ are given by the projection $Q^L_{MK}$. 

As above, there are selection rules given by the symmetry. In contrast to the previous cases, the states $\ket{{}^2\Psi}$ defined above are not axially symmetric and therefore projections with $K\neq 0$ are possible. The remaining symmetries of these states are 
\begin{eqnarray}\label{eq:N3sym}
  U(0,0,\pi)\ket{{}^2\Psi}&=&\ket{{}^2\Psi}, \nonumber\\
  U(0,\pi,0)\ket{{}^2\Psi}&=&-\ket{{}^2\Psi}, \\
  SU(0,0,\pi/2)\ket{{}^2\Psi}&=&\ket{{}^2\Psi}. \nonumber
\end{eqnarray}
where the arguments of the rotation operator $U(\phi,\theta,\gamma)$ are the three Euler angles. The last symmetry involves the particle-hole transformation $S$. This symmetry is special to \csixtythreeminus{} where the number of holes is equal to the number of electrons. The operator $S$ is defined by
\begin{eqnarray}
  Sc^\dagger_{ns}S^\dagger =c_{n-s}, &&
  Sc_{ns}S^\dagger         =c^\dagger_{n-s},\\
  Sa^\dagger_kS^\dagger    = -a_k, &&
  Sa_kS^\dagger            =-a^\dagger_k. \nonumber
\end{eqnarray}
The definition implies that $S$ commutes with the angular momentum, i.e. 
$[\V L,S]=0$, and therefore $[U(\Theta),S]=[Q_{MK}^L,S]=0$. Selection rules for the quantum numbers $LK$ can be derived using the symmetries~(\ref{eq:N3sym}) and the following properties of the projection operators
\begin{eqnarray}\label{eq:Qprops}
  Q^L_{MK}\,U(0,0,\pi)&=&(-1)^K\, Q^L_{MK}, \nonumber\\
  Q^L_{MK}\,U(0,\pi,0)&=&\sigma_K\,(-1)^L\, Q^L_{MK}, \\
  Q^L_{MK}\,S\,U(0,0,\pi/2)&=&(-1)^{\frac{K}{2}}\,S\, Q^L_{MK} \; \textrm{for $K$ even}. \nonumber
\end{eqnarray}
The first property implies that $K$ is even. The second property implies that $K\geq0$ for odd $L$ and $K<0$ for even $L$ due to the definition~(\ref{eq:sigmadef}) of $\sigma_K$. Note that $K=0$ states are forbidden for even $L$. These rules agree with the literature~\cite{art:obrien96}. The last property, applied in the second equality below, yields the rule for the orthogonality of projections:
\begin{eqnarray}\label{eq:projortho}
  \bra{{}^2\Psi}Q^{L_2}_{K_2M_2}Q^{L_1}_{M_1K_1}\ket{{}^2\Psi}=
  \delta_{L_1L_2}\delta_{M_1M_2}
    \bra{{}^2\Psi}Q^{L_1}_{K_2K_1}\ket{{}^2\Psi} \nonumber \\
    =(-1)^{\frac{K_1+K_2}{2}}\delta_{L_1L_2}\delta_{M_1M_2}
    \bra{{}^2\Psi}Q^{L_1}_{K_2K_1}\ket{{}^2\Psi}.\quad
\end{eqnarray} 
This implies that projections with equal $L$ and $M$ are only orthogonal if $(K_1+K_2)/2$ is odd. With these selection rules and the spectrum of the top given in~(\ref{eq:n3asym}), we find that the ground state and lowest excitations have quantum numbers $(LK)=(10),(2-\!2),(4-\!4),(32),(30)$. The next higher state in energy is $(6-\!6)$ which, according to~(\ref{eq:projortho}), is allowed to mix with $(6-\!2)$.   

In order to calculate rotator excitations, the basis states~(\ref{eq:c60m3statbasis}) are projected:
\begin{equation}\label{eq:c60m3dynbasis}
  \ket{{}^2X_P^K}=Q^L_{0K}\ket{{}^2\Psi_P},\quad
  \ket{{}^2X_D^K}=Q^L_{0K}\ket{{}^2\Psi_P}.
\end{equation}  
Again, $X$ denotes the total angular momentum and $K$ is the quantum number for the rotation around the principal axis of the top. These two states are eigenstates of $H_J$ with eigenvalues $2J$ and $0$ respectively.
The calculation of the expectation value of $H$ proceeds as before. The matrix elements are given by:
\begin{eqnarray}
   \frac{\bra{{}^2X_P^K} H \ket{{}^2X_P^K}}{\langle {{}^2X_P^K} \ket{{}^2X_P^K}}&=&
   \frac{5}{2} +\frac{q^2}{2}\frac{G^P_{LK}}{N^P_{LK}}+\frac{2J}{\bar\omega},
      \nonumber \\
   \frac{\bra{{}^2X_D^K} H \ket{{}^2X_D^K}}{\langle {{}^2X_D^K} \ket{{}^2X_D^K}}&=&
   \frac{5}{2} +\frac{q^2}{2}\frac{G^D_{LK}}{N^D_{LK}},\\
   \frac{\bra{{}^2X_D^K} H \ket{{}^2X_S^K}} 
      {\sqrt{\langle {{}^2X_S^K} \ket{{}^2X_S^K}\langle {{}^2X_D^K} \ket{{}^2X_D^K}}}&=&
      \frac{\sqrt 3}{2}\,gq\,\frac{N^P_{LK}+N^D_{LK}}{
	        \sqrt{N^D_{LK}N^P_{LK}}}, \nonumber
\end{eqnarray}
where
\begin{eqnarray}
  N^P_{LK}(q)&=&\frac{2l\!+\!1}{8\pi^2}\int d\Theta\, 
    D^L_{KK}D^1_{00}e^{-\frac{q^2}{2}(1-D^2_{22})},\nonumber \\
  N^D_{LK}(q)&=&\frac{2l\!+\!1}{8\pi^2}\int d\Theta\, 
    D^L_{KK}D^2_{-2-2}e^{-\frac{q^2}{2}(1-D^2_{22})}, \\
  G^P_{LK}(q)&=&\frac{2l\!+\!1}{8\pi^2}\int d\Theta\, 
    D^L_{KK}D^1_{00}D^2_{22}e^{-\frac{q^2}{2}(1-D^2_{22})},\nonumber \\
  G^D_{LK}(q)&=&\frac{2l\!+\!1}{8\pi^2}\int d\Theta\, 
    D^L_{KK}D^2_{-2-2}D^2_{22}e^{-\frac{q^2}{2}(1-D^2_{22})}.\nonumber 
\end{eqnarray}
In the expressions above, the integration over the Euler angles $\phi$, $\gamma$ is not trivial. As can be seen from~(\ref{eq:rwd}), the Wigner D-functions $D^L_{KK}$ depend on $\phi$,$\gamma$ through $\cos[K(\phi\pm\gamma)]$. Therefore, the integration over $\phi$, $\gamma$ can be carried out in terms of modified Bessel functions $I_n$ using the new integration variables $\mu=\phi+\gamma$ and $\nu=\phi-\gamma$. For example, $N^P_{10}$ becomes
\begin{eqnarray}
  N^P_{10} =
  \frac{3}{2}\int_{-1}^1\! dt\, t^2\,e^{-\frac{q^2}{2}}\,
    I_0\!\left[\frac{q^2(1+t)^2}{8}\right]I_0\!\left[\frac{q^2(1-t)^2}{8}\right].
\end{eqnarray} 
The remaining integrals are numerically evaluated. The lower eigenvalue is then minimized with respect to $q$ in order to find the variational ground state energy.
Fig.~\ref{fig:c60m3nrg} shows the resulting energies for $J=0$ and $J=\bar\omega$.  
The asymptotic spectrum~(\ref{eq:n3asym}) is also shown in Fig.~\ref{fig:c60m3nrg} and differs from the energies of the present calculation. Again, this is improved by adding two more states, $A_{ep}^\dagger\ket{{}^2X_P^K}$ and $A_{ep}^\dagger\ket{{}^2X_D^K}$, to the Hilbert space (see Fig.~\ref{fig:c60m3nrg})~\cite{note:thesispaper1}.

As discussed above, a level crossing from a low-spin to a high-spin state occurs in \csixtythreeminus{} for large enough $J$. The corresponding line in the $(g,J)$
parameter space is shown in Fig.~\ref{fig:c60xing} and was calculated using the improved variational approach. The criterion $J/(\bar\omega g^2)=0.381$ derived above becomes correct in the large $g$ limit. For $g\to 0$, the line ends at $J/(\bar\omega g^2)=3/4$ which can be shown using perturbative results~\cite{art:auerbach94} for small $g$.

\section{Results for \csixty}\label{sec:c60spec}

\begin{table}
\begin{center}
\begin{tabular}{c|c|cc|ccc}
  && \multicolumn{2}{c|}{$J=0$}&\multicolumn{3}{c}{$J=\bar\omega$}\\
  N&& Ref.~\onlinecite{thesis:Manini} & Present & 
    & $\partial_g$ & $\partial_J$ \\
  \hline\hline  
  1&$E(P)$ & $-139.6$ & $-132.8$ & $-132.8$ & $-137.6$ & 0 \\ \cline{2-7}
  &$E(F)-E(P)$ & 26.8 & 40.9 & 40.9 & $-20.6$ &  0 \\ 
  &$E(H)-E(P)$ & 53.9 & 87.4 & 87.4& $-37.4$ & 0 \\
  &$E(L)-E(P)$ & - & 137.3 & 137.3& $-52.0$ & 0 \\
  \cline{2-7}
  & $U'$ & $-126.7$ & $-141.2$ & $-5.6$ & $-167$ & 1.75 \\
  \hline\hline
  2&$E({}^1\!S)$ &      $-405.9$&$-406.8$ & $-271.2$ & $-442.2$ & 1.75  \\ \cline{2-7}
  &$E({}^1\!D)\!-\!E({}^1\!S)$ &13.2&  9.8 & 9.6 & $-9.8$& 0.003 \\
  &$E({}^1\!G)\!-\!E({}^1\!S)$ &38.4&  29.5 & 29.4 & $-31.4$ & 0.010\\
  &$E({}^1\!I)\!-\!E({}^1\!S)$ &74.8&  56.0 & 56.2 & $-55.6$&  0.018 \\ \cline{2-7}
  &$E({}^3\!P)\!-\!E({}^1\!S)$ & 266.3 & 273.9 & 66.8 & 304.6 & -2.75  \\
  \cline{2-7}
  & $U'$ & 329.3 & 345.1 & 141.6 & 386.5 & -2.63 \\
  \hline\hline
  3&$E({}^2\!P^0)$ & $-342.9$&$-335.7$ & $-268.0$ & $-360.3$ & 0.87  \\ \cline{2-7}
  &$E({}^2\!D^{\bar 2})\!-\!E({}^2\!P^0)$ &-&  5.0 & 4.8 & $-4.2$& -0.002  \\
  &$E({}^2\!G^{\bar 4})\!-\!E({}^2\!P^0)$ &-&  28.1 & 28.0 & $-22.6$ & 0.002\\
  &$E({}^2\!F^2)\!-\!E({}^2\!P^0)$ &-&  35.4 & 35.6 & $-24.8$&  0.007 \\ \cline{2-7}
  &$E({}^4\!S)\!-\!E({}^2\!P^0)$ & 342.9 & 335.7 & 51.7 & 360.3 & -3.87 \\
  \cline{2-7}
  & $U'$ & $-126.0$ & $-142.2$ & $-6.4$ & $-163.8$ & 1.76 
\end{tabular}
\caption{\label{tab:lowstates}
Ground state energies and low-energy excitations for \csixtynminus{} ($N=1,2,3$) as calculated by the effective mode approximation (EMA). $U'(N)=E_0(N-1)+E_0(N+1)-2E_0(N)$ is the contribution of the Jahn-Teller and Hund's rule coupling Hamiltonian $H'=H_p+H_{ep}+H_J$ to the effective on-site repulsion~\cite{art:gunnarsson95b} $U$. Improved variational states with $g=1.532$, $\bar\omega=72.1$~meV, $J=0$ (columns 3,4) and $J=\bar\omega$ (columns 5-7) are used. All energies are in meV. The zero-point energy $\omega_0=\sum_\alpha (5/2) \omega_\alpha$ is neglected. Column 3 is the exact diagonalization result from Ref~\onlinecite{thesis:Manini}. Columns 6,7 are the derivatives with respect $g$ and $J$. }
\end{center}
\end{table}

In this section we calculate ground state and excitation energies for parameters specific to \csixty{}. As shown in section~\ref{sec:ham}, the ground state energy in the effective mode approximation (EMA) is a variational estimate for the ground state energy of the full multi-mode model. Ground state energies and lowest excitations obtained from the EMA using improved variational approaches are given in Tab.~\ref{tab:lowstates}. Concerning the phonon-related parameters $g$ and $\bar\omega$, there is a consensus in the literature to use the parameter set given in Tab.~\ref{tab:epparameters} which originates from photoemission experiments on \csixtyminus{} in gas phase~\cite{art:gunnarsson95b}. This parameter set yields $g=1.532$ and $\bar\omega=72.1$~meV for the EMA. On the other hand, there is much less consensus concerning the Hund's rule coupling $J$. Below, we will first discuss the case $J=0$ where exact diagonalization results are available~\cite{thesis:Manini}. Subsequently, we will determine the parameter $J>0$ through the singlet-triplet gap in \csixtytwominus{} which can be measured experimentally. 

We first consider the case $J=0$. As can be seen from Tab.~\ref{tab:lowstates}, the relative errors between the present ground state energies and the exact diagonalization results~\cite{thesis:Manini} are 4.9\%, -0.2\%, 2.1\% for $N=1,2,3$ respectively. Hence, the present results agree well which confirms the validity of the EMA. The agreement is best for $N=2$ where the variational energy is in fact below the exact diagonalization result. This is due to the truncation of the phonon Hilbert space in the exact diagonalization approach which makes it variational as well. Hence, exact diagonalization tends to overestimate the ground state energy for a large number of excited phonons which is the case for $N=2$ where the Jahn-Teller energy gain is largest. The present approach based on coherent phonon states doesn't suffer from this truncation. The lowest excitations are also given in Tab.~\ref{tab:lowstates}. For $N=1$, the exact diagonalization yields lower energies than the present approach opposite to $N=2$. This is again due to the truncation effect discussed above. In addition, the exact diagonalization yields also the vibronic excitations which are not captured by the present approach. As discussed in Ref.~\onlinecite{thesis:Manini}, the two lowest levels for $N=1$ are the $L=1,3$ rotator states whereas the third level are $L=2$ vibronic excitations~\cite{thesis:Manini}. In the case $N=2$, the $L=0,2,4,6$ rotator states are lowest in energy, followed again by $L=2$ vibronic excitations~\cite{thesis:Manini}. Ref.~\onlinecite{thesis:Manini} doesn't provide excitation energies for $N=3$. However, as can be seen from Tab.~\ref{tab:lowstates}, the low-energy rotator excitation for $N=3$ are smaller than those for $N=2$. This suggests that the four lowest levels of $N=3$ are also pure rotator states. 

\begin{figure}
\begin{center}
\includegraphics[width=0.45\textwidth]{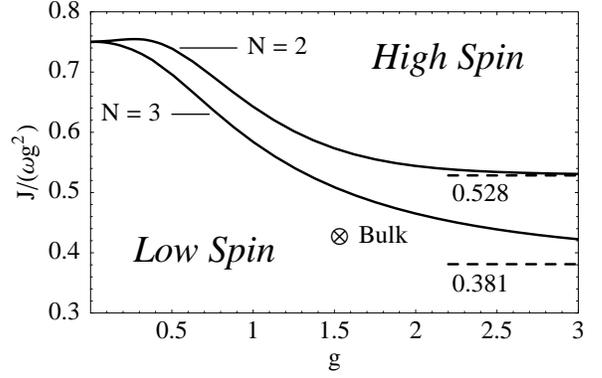}
\caption{\label{fig:c60xing}
  Boundaries between the low-spin and high-spin sectors in the $(g,J)$ parameter space. The upper line is for $N=2$ and indicates the ${}^1S$-${}^3P$ level crossing whereas the lower line is for $N=3$ indicating the ${}^2P^0$-${}^4S$ level crossing. Both lines end at $J/(\bar\omega g^2)=3/4$ for $g\to 0$. For $g\to\infty$, the upper line approaches the constant value 0.5284 whereas the lower line approaches $3(4-\sqrt{15})=0.381$. As discussed in the text, realistic parameters for bulk \csixty{} are $g=1.532$ and $J=\bar\omega=72.1$~meV which is indicated by the cross.
    }
\end{center}
\end{figure}
In the literature, estimations for the value of $J$ differ largely. Theoretical values range between $J=15-300$~meV (see Ref.~\onlinecite{art:wierzbowska04} and references therein). Experimentally, $J$ is not directly accessible. However, the low-spin/high-spin gap for $N=2,3$ can be measured by various means. As can be seen in the last column of Tab.~\ref{tab:lowstates}, this gap depends strongly on $J$. Below we determine the value of $J$ using experimental values for the low-spin/high-spin gap. There is a consensus that isolated \csixtynminus{} ($N=2,3$) ions are in the low-spin state (see Ref.~\onlinecite{art:reed0} and references therein). However, there was a controversy~\cite{art:reed0} on whether the low-spin/high-spin gap is very small (below one wavenumber~\cite{art:boyd95}) or rather of the order of 600 wavenumbers~\cite{art:trulove95}. Recently, this problem was carefully reconsidered and it was shown that activated behaviors of \csixtynminus{} which were observed so far and used to determine the gap are in fact due to C$_{120}$O impurities~\cite{art:paul02,art:drew03}. However, the work clearly reconfirms that isolated \csixtynminus{} ions ($N=2,3$) are in the low-spin state. In view of Fig.~\ref{fig:c60xing} this implies that $J/(g^2\bar\omega)<0.5$ which yields the upper bound $J<85$~meV for isolated \csixtynminus{}. Measurements of the low-spin/high-spin gap exist for \csixty{} bulk materials, in particular for \kfourcsixty{} which is a non-magnetic insulator. The \csixtyfourminus{} ion in this material is equivalent to the \csixtytwominus{} ion by particle-hole symmetry. Hence, the low-spin/high-spin gap should correspond to the singlet-triplet gap $\Delta_{ST}$ of \csixtytwominus{}. This gap is observed in  magnetic susceptibility~\cite{art:lukyanchuk95} and spin relaxation~\cite{art:zimmer94,art:zimmer95,art:kerkoud96,art:brouet02} measurements on \kfourcsixty{}. The magnetic susceptibility and the spin relaxation scale with the thermal occupation of the triplet
${}^3P$ state which shows an activated behavior with a gap
$\Delta_{SP}\approx 50-100$~meV. Using the $J$-dependence of the gap as given in Tab.~\ref{tab:lowstates}, one deduces $J\approx 60-80$~meV in agreement with the upper bound 85~meV found above. For convenience we suggest $J=\bar\omega=72.1$~meV which leads to low-spin/high-spin gaps of 66.8~meV and 51.7~meV for $N=2,3$ respectively. Energies for $J=\bar\omega$ are given in Tab.~\ref{tab:lowstates} together with derivatives with respect to $g$ and $J$. The $J$-dependence in the low-spin sector of $N=2,3$ is non-trivial. However, to a good approximation, levels are shifted linearly and in parallel for a given $N$. Therefore, excitation energies in the low-spin sectors depend little on $J$ as is confirmed in Tab.~\ref{tab:lowstates}. 

Comparing the ground state energies for the cases $J=0$ and $J=\bar\omega$ in Tab.~\ref{tab:lowstates} shows that the Jahn-Teller effect is partly counterbalanced by the Hund's rule coupling. This observation is particularly relevant for the corrections to the on-site repulsion $U$. Generally, the main contribution to the effective on-site repulsion is the isotropic Coulomb repulsion $H_U$ which we separated from the Hund's rule coupling in Hamiltonian~(\ref{eq:H}). In addition, there is a second contribution~\cite{art:gunnarsson95b,art:gunnarsson97} $U'(N)=E_0(N-1)+E_0(N+1)-2E(_0N)$ which is due to the different ground state energies of $H'=H_p+H_{ep}+H_J$ for different occupation numbers $N$. It was argued that this contribution is not negligible and may explain why compounds with average occupation number $N=2,4$ are insulating whereas compounds with $N=3$ are mostly metallic. Indeed, for $J=0$ we have $U'(2)-U'(3)\approx 0.5$~eV which is important compared to $U=1-2$~eV. For $J=\bar\omega$, the difference $U'(2)-U'(3)$ is reduced to 0.14~eV which is an order of magnitude smaller than $U$. Hence, including $J$ reduces $U'$ significantly.

\section{Conclusion}\label{sec:conclusion}

With the present approach, variational wavefunctions for the ground state and rotator excitations of \csixtynminus{} ions are constructed semi-analytically. The Jahn-Teller physics, where the EMA is used, and the Hund's rule coupling are treated on the same level. The strict use of the SO(3) symmetry and projection operators allows for an efficient formalism. In this formalism it is evident that any scalar operator, such as the Hamiltonian itself, commutes with any projection operator. Thanks to this property, only one integration has to be done numerically in the final expressions of the expectation values. This is a major achievement over previous approaches and enables the present approach to go beyond previous results.

In a first step we calculate ground state energy and rotator excitations for the three distinct cases $N=1,2,3$ whereby each case has some additional complication compared to the previous one. The simplest case is $N=1$ and served to explain the projection operator technique in detail. We find low-energy excitation in agreement with previous works. The correct asymptotic behavior is recovered when the improved version with an enlarged Hilbert space is used. The main challenge for $N=2$ is the additional Hund's rule coupling. We investigated the competition of Jahn-Teller effect and Hund's rule coupling on the level of both, projected and unprojected states. We find, somewhat in contradiction to the general picture, that strong Hund's coupling doesn't completely suppress the Jahn-Teller effect, but rather reduces the effective electron-phonon coupling constant by a factor of 2 within the $S=0$ sector. Of course, strong enough Hund's rule coupling favors the $S=1$ state. We calculate the separation between the low-spin and high-spin sector in the complete $(g,J)$ parameter space. The difficulty of $N=3$ lies in the fact that the unprojected state minimizing the electron-phonon coupling is not anymore axially symmetric. The problem therefore becomes similar to a symmetric top and states involve a third quantum number $K$. Using the symmetries of the unprojected state we deduce the allowed values for the quantum numbers $LK$ in agreement with previous findings. A new result is, that two states with $L_1=L_2$ and odd $(K_1+K_2)/2$ are allowed to mix. The evaluation of matrix elements for $N=3$ is more complicated and involves Bessel functions. Nevertheless, only one numerical integration is required.  

Using the results of the previous sections, we calculated ground state energy and lowest excitations in agreement with exact diagonalization results. In addition, we give a thorough discussion of the parameters specific to \csixty{}. Whereas there is a consensus on the value of the electron-phonon coupling, there is much uncertainty on what concerns the Hund's rule coupling. We use the present results to make a connection between the Hund's rule coupling constant and the low-spin/high-spin gap which is experimentally accessible. This allows us to pin down the Hund's rule coupling constant to $J=60-80$~meV. Using $J=\bar\omega=72.1$~meV to calculate the ground state energies, we find that the finite Hund's rule coupling partly counterbalances the Jahn-Teller energy gain and that the ground state energies for the cases $N=2,3,4$ become almost equal. Therefore, the contribution to the on-site repulsion arising from the Jahn-Teller effect is substantially reduced when including the Hund's rule coupling.

The authors thank T.M. Rice, C. Helm and I. Milat for fruitful discussions. This work has been supported by the Swiss Nationalfonds, by the NCCR MaNEP and by the Center for Theoretical Studies of ETH Zurich. 

\appendix
\section{Real representation}\label{app:rr}

\begin{table}
\begin{center}
\begin{tabular}{c||c|c||c|c}
  Symmetry:   & SO(3) & $I_h$ & SO(3) & $I_h$ \\
  \hline	
  IR: & $L=1$ & $\;\;t_{1u}\;\;$ & $L=2$ & $H_g$ \\
  \hline
     & $M=-1$ & $y$ & $M=-2$ & $z$ \\ 
     & $M=0$ & $z$ & $M=-1$ & $x$ \\ 
     & $M=1$ & $x$ & $M=0$ & $\sqrt{3/8}\theta - \sqrt{5/8}\varepsilon$ \\ 
     & & & $M=1$ & $y$ \\ 
     & & & $M=2$ & $\sqrt{5/8}\theta + \sqrt{3/8}\varepsilon$ \\ 
  \end{tabular}
  \caption{\label{tab:IRcomponents}
Relationship between the components of the $L=1,2$ IR's of SO(3) as used in the present work (columns 2 and 4) and the components of the $t_{1u}$ and $H_g$ IR's of the icosahedral symmetry $I_h$ as defined in Ref.~\onlinecite{art:fowler85} (columns 3 and 5). Note that the notation used in columns 3 and 5, i.e. the letters $x,y,z,\theta,\varepsilon$ denoting the different components, follows Ref.~\onlinecite{art:fowler85}.
    }
\end{center}
\end{table}
In the following we discuss the transformation to real spherical harmonics. This transformation applies not only to spherical
harmonics, but to all quantities depending on the angular momentum
quantum numbers $(lm)$, such as the Wigner D-functions and the
Clebsch-Gordan coefficients. For the purpose of a clear notation, quantities in
the complex spherical harmonics basis will be written with a
tilde. The transformation from complex spherical harmonics $\tilde Y_{lm}(\Omega)$ to real spherical harmonics $Y_{ln}(\Omega)$ is defined by
\begin{equation}\label{eq:defrsh}
    Y_{ln}(\Omega)=\sum_{m=-l}^l \Lambda_{nm}\,\tilde Y_{lm}(\Omega),
\end{equation}
where
\begin{equation}
    \Lambda_{nm}=\sqrt{\sigma_n}\,[\delta_{nm}+ \sigma_n\delta_{n-m}]\,\beta_m,
\end{equation}
with
\begin{equation}\label{eq:sigmadef}
  \sigma_n=\left\{\begin{array}{cc}
      1, & n\geq0, \\
      -1, & n < 0, 
      \end{array}\right. \;
  \beta_m=\left\{\begin{array}{cc}
      (-1)^m\frac{1}{\sqrt 2}, & m>0, \\
      \frac{1}{2}, & m = 0, \\
      \frac{1}{\sqrt 2}, &  m < 0.
      \end{array} \right.
\end{equation}
The coefficients $\Lambda_{nm}$ are best represented as a matrix. For $-2\ge n,m \ge 2$, $\Lambda$ takes the form
\begin{equation}
  \Lambda=\left( \begin{array}{ccccc}
      \frac{i}{\sqrt 2} &0&0&0& -\frac{i}{\sqrt 2} \\
      0 & \frac{i}{\sqrt 2} & 0 & \frac{i}{\sqrt 2} & 0 \\
      0 & 0 & 1 & 0 & 0 \\
      0 & \frac{1}{\sqrt 2} & 0 & -\frac{1}{\sqrt 2} & 0 \\
      \frac{1}{\sqrt 2} &0&0&0& \frac{1}{\sqrt 2}
      \end{array} \right).
\end{equation}
With definition~(\ref{eq:defrsh}), the $\phi$-dependence of $Y_{ln}(\theta,\phi)$ is $\cos(n\phi)$ for $n\geq0$ and $\sin(n\phi)$ for negative $n<0$. Creation and annihilation operators are transformed in the same way:
\begin{equation}\label{eq:relops}
  c_{ns}^\dagger = \sum_{m=-1}^1 \Lambda_{nm}\, \tilde c_{ms}^\dagger , \quad
  c_{ns} = \sum_{m=-1}^1 \Lambda_{nm}\, (-1)^m \, \tilde c_{-ms}.
\end{equation}
The operators $a_{k}^\dagger$ and $a_{k}$ are given by the same rules. Note that $\tilde c_{ms}^\dagger$ creates an electron such that $L_z\,\tilde c_{ms}^\dagger\ket 0=m\,\tilde c_{ms}^\dagger\ket 0$ whereas the electrons created by $c_{ns}^\dagger$ have $y$, $z$ or $x$ symmetries for $n=-1,0,1$.
Using the definition $\tilde D_{mk}^l(\Theta)=\bra{lm}U(\Theta)\ket{lk}$ for the complex Wigner D-functions, the following transformation rule to the real Wigner D-functions $D^L_{MK}$ can be deduced
\begin{equation}\label{eq:ap.trans.D.rule}
    D^L_{MK}(\Theta)=\sum_{mk}\Lambda_{Mm}^*\,\Lambda_{Kk}\,\tilde D^L_{mk}(\Theta).
\end{equation}
The real Wigner-D functions are explicitly given by
\begin{eqnarray}\label{eq:rwd}
  \lefteqn{D^L_{MK}(\phi,\theta,\gamma)=
     2 \beta_M \beta_K d_{MK}^L(\theta) \cos(M\phi+K\gamma)+} \\
     &&2 \sigma_M \beta_M \beta_{-K} d_{M-K}^L(\theta) \cos(M\phi-K\gamma)
  \;\;\textrm{if}\;\;\sigma_M \sigma_K = 1, \nonumber  \\
  \lefteqn{D^L_{MK}(\phi,\theta,\gamma)=
     2 \sigma_M \beta_M \beta_K d_{MK}^L(\theta) \sin(M\phi+K\gamma)-}\nonumber\\
     && 2\beta_M \beta_{-K} d_{M-K}^L(\theta) \sin(M\phi-K\gamma) 
  \;\;\textrm{if}\;\;\sigma_M \sigma_K = -1.  \nonumber
\end{eqnarray}
The functions $d_{MK}^L(\theta)$ are the same as used for the complex Wigner-D functions 
$\tilde D^l_{mk}=e^{-i\phi m}\,d^l_{mk}(\theta)\,e^{-i\gamma k}$. They are tabulated in various references~\cite{book:Butler,book:QToAngularMomentum}.
The real Wigner-D functions describe the rotation of tensor operators $T_{LM}$, such as $c_{ns}$, $c_{ns}^\dagger$, $a_{k}$ and $a_{k}^\dagger$, which have the symmetries of the real spherical harmonics: 
\begin{equation}\label{eq:Trot}
  U(\Theta) T_{LM} U^\dagger(\Theta)=
     \sum_{M'=-L}^L D^L_{M'M}(\Theta) T_{LM}
\end{equation}
Finally, the Clebsch-Gordan coefficients for the real spherical harmonics are given by
\begin{equation}\label{eq:rclebschgordan}
    R^{LM}_{L_1M_1\,L_2M_2}=\sum_{m m_1 m_2}
    \Lambda_{M_1m_1}^* \, \Lambda_{M_2m_2}^* \, \Lambda_{Mm} \,
    C^{Lm}_{L_1m_1\,L_2m_2},
\end{equation}
where $C^{Lm}_{L_1m_1\,L_2m_2}$ are the usual Clebsch-Gordan coefficients for the complex spherical harmonics. The Clebsch-Gordan coefficients $R^{LM}_{L_1M_1\,L_2M_2}$ have the property
\begin{equation}
  R^{*\,LM}_{L_1M_1\,L_2M_2}=R^{LM}_{L_2M_2\,L_1M_1}=
  (-1)^{L+L_1+L_2}R^{LM}_{L_1M_1\,L_2M_2}
\end{equation}
which implies that they are real if $L+L_1+L_2$ is even and imaginary otherwise. Furthermore the following orthogonality relation holds
\begin{equation}
  \sum_{LM}R^{LM}_{L_1M_1\,L_2M_2}\,R^{LM}_{L_2M_2'\,L_1M_1'}=
  \delta_{M_1M_1'}\,\delta_{M_2M_2'}.
\end{equation}
The present work is formulated in terms of the SO(3) symmetry where the $p\otimes d$ electron-phonon coupling involves the Clebsch-Gordan coefficients $R^{2M}_{1M_1\,1M_2}$. As discussed in the introduction, this is equivalent to the $t_{1u}\otimes H_g$ electron-phonon coupling in the context of icosahedral symmetry. The icosahedral formulation involves the icosahedral Clebsch-Gordan Coefficients tabulated in Ref.~\onlinecite{art:fowler85}. The two sets of Clebsch-Gordan coefficients are equal when using the relationships between the components of the IR's as given in Tab.~\ref{tab:IRcomponents}


\begin{thebibliography}{100}
\bibitem{note:thesispaper1}
All calculation were performed with Mathematica.
\end{thebibliography}
\end{document}